\documentclass[letterpaper,11pt]{article}
\pdfoutput=1 

\usepackage{jheppub} 

\usepackage{tikz}

\def \Tr{\mbox{Tr\,}}

\def \sgn{\mbox{sgn\,}}

\title{Bi-Local Holography in the SYK Model}

\author{Antal Jevicki,}
\author{Kenta Suzuki,}
\author{Junggi Yoon}

\affiliation{Department of Physics, Brown University,\\
182 Hope Street, Providence, RI 02912, U.S.A.}

\emailAdd{antal\_jevicki@brown.edu}\emailAdd{kenta\_suzuki@brown.edu}
\emailAdd{jung-gi\_yoon@brown.edu}

\preprint{{\tt BROWN-HET-1673}}

\abstract{
We discuss large $N$ rules of the Sachdev-Ye-Kitaev model and the bi-local representation of holography of this theory.
This is done by establishing $1/N$ Feynman rules in terms of bi-local propagators and vertices, which can be evaluated following 
the recent procedure of Polchinski and Rosenhaus. These rules can be interpreted as Witten type diagrams of the dual AdS theory,
which we are able to define at IR fixed point and off.}

\begin{document}
\maketitle
\flushbottom

\section{Introduction}
\label{sec:intro}

Simple models of AdS/CFT duality are valuable laboratories for in-depth understanding of holography and quantum features of black holes. 
One such model is the Sachdev-Ye-Kitaev (SYK) model~\cite{Kitaev:2015,Kitaev:2014,Sachdev:2015efa,Polchinski:2016xgd} arising recently from the earlier Sachdev-Ye (SY) model \cite{Sachdev:1992fk,Georges:1999,Sachdev:2010um,Sachdev:2010uj} representing a system of one dimensional fermions quantized with quenched disorder.
A similar problem is studied in \cite{Erdmenger:2015xpq}.
This model can be studied in the $1/N$ expansion and at strong coupling with an IR fixed point and potentially very interesting AdS$_2$ dual. 
Kitaev~\cite{Kitaev:2015}, has demonstrated the chaotic behavior of the strongly coupled system (at Large $N$)
in terms of the Lyapunov exponent and has exhibited some elements of the dual black hole.
 Recently, Polchinski and Rosenhaus~\cite{Polchinski:2016xgd} have evaluated the spectrum of two particle states of the SYK model
by solving the large-$N$ Schwinger-Dyson (SD) equations of the four-point function. 
Their solution (and earlier works \cite{Kitaev:2015,Sachdev:2015efa}) employ the emergent conformal symmetry of the model and correspondingly exhibit characteristic features of the AdS theory. Further the in-depth studies \cite{Maldacena:2016,Maldacena:2016hyu,Fu:2016yrv} of the model are of definite interest specially in view of the appearance of the thermal butterfly effect~\cite{Shenker:2013pqa,Shenker:2013yza,Leichenauer:2014nxa,Roberts:2014isa,Roberts:2014ifa,Shenker:2014cwa,Maldacena:2015waa,Polchinski:2015cea,Stanford:2015owe,Michel:2016kwn,Caputa:2016tgt,Gu:2016hoy,Perlmutter:2016pkf,Anninos:2016szt,Turiaci:2016cvo} characteristic of black holes.
In this paper, inspired by these positive results of \cite{Polchinski:2016xgd}, we discuss a description of the SYK theory in terms of explicit
$1/N$ rules given in terms of bi-local propagators and vertices \cite{Das:2003vw} associated with $O(N)$ symmetry appearing after the quenched disorder averaging. This collective representation systematically incorporates arbitrary $n$-point bi-local correlators and generates the corresponding SD equations in terms of closed set of observables that provide a holographic interpretation \cite{Das:2003vw}. 
For the generic case of $N$-component Vector/Higher Spin Gravity duality \cite{Klebanov:2002ja,Sezgin:2002rt}
bi-locality provides an explicit implementation of holography \cite{Das:2003vw,Koch:2010cy,Koch:2014mxa,Koch:2014aqa,Mintun:2014gua,Brodsky:2014yha}. The present SYK system shares analogous features and can be studied along the same line, as the $d=1$ case with an AdS$_2$ dual.
The collective fields (bi-local or loop) can be also termed as `precursors' \cite {Polchinski:1999yd} and their ability to encode bulk information relates to the completeness of these variables at Large $N$. We also mention a similarity with another $c=1$/$2D$ duality representing matrix mechanics and non-critical string theory.
The overview of our work is as follows: After reviewing the basic features of the SYK model in the replica formulation we formulate in Section~\ref{sec:replica collective field theory} the corresponding bi-local collective theory with explicit $1/N$ rules given in terms of propagators and vertices.
We use this to consider the $n\to 0$ limiting theory (with $n$ being the replica number) which is a necessary step in transition to the dual description. 
In Section~\ref{sec:ads bulk}, we present elements of the limiting theory. The bi-local propagator defining the $1/N$ diagrams is evaluated in the basis established in \cite{Polchinski:2016xgd}. It is seen to lead the same poles as in \cite{Polchinski:2016xgd} with a (non-polynomial) AdS bulk representation.
We show that through introduction of a collective coordinate for time, we eliminate the zero mode problem and are able to define the theory in the IR limit.
The cubic vertex is presented in detail, and analogous higher vertices are seen to be specified, which can be interpreted as Witten type rules of the bulk theory.
In Appendices~\ref{app:higher loop diagrams} and \ref{app: two-point function}, it is seen that these vertices lead to correct two ( i.e. 4 \footnote{ in terms of the fundamental fermions}) and three ( i.e. 6 ) point large-$N$ Green's functions.

\section{Replica Collective Field Theory}
\label{sec:replica collective field theory}

In the series of talk \cite{Kitaev:2015}, Kitaev presented a simplified version of the Sachdev-Ye model, which nevertheless exhibits holographic features and chaotic behavior.
This $N$-site fermionic theory is represented by the Hamiltonian
	\begin{equation}
		H \, = \, \frac{1}{4!} \sum_{i,j,k,l=1}^N J_{ijkl} \, \chi_i \, \chi_j \, \chi_k \, \chi_l \, ,
	\end{equation}
where $\chi_i$ are Majorana fermions, which satisfy $\{ \chi_i, \chi_j \} = \delta_{ij}$.
The Lagrangian of this model is given by
	\begin{equation}
		L \, = \, - \, \frac{1}{2} \sum_{i=1}^N \chi_i \partial_t \chi_i \, - \, \frac{1}{4!} \sum_{i,j,k,l=1}^N J_{ijkl} \, \chi_i \, \chi_j \, \chi_k \, \chi_l \, .
	\end{equation}
with Euclidean time $t$.
The coupling constant $J_{ijkl}$ are random and the disordered probability is specified by 
	\begin{equation}
		P(J_{ijkl}) \, = \, \sqrt{\frac{N^3}{12 \pi J^2}} \exp\left( - \, \frac{N^3J_{ijkl}^2}{12J^2} \right) \, .
	\end{equation}
The replica method \cite{Sachdev:2015efa,Anninos:2016szt} reduces the problem to the evaluation of
	\begin{equation}
		\langle \ln Z \rangle_J \, = \, \lim_{n \to 0} \left( \frac{\langle Z^n \rangle_J - 1}{n} \right) \, ,
	\label{replica trick}
	\end{equation}
where the disordered average is defined by
	\begin{equation}
		\langle \mathcal{O} \rangle_J \, \equiv \, \int \prod_{i,j,k,l}^N dJ_{ijkl} \, P(J_{ijkl}) \, \mathcal{O} \, .
	\end{equation}
The disordered average over the couplings $J_{ijkl}$ in the replica representation is readily evaluated with the result \cite{Sachdev:2015efa}:
	\begin{equation}
		\langle Z^n \rangle_J
		\, = \, \int \mathcal{D}\chi_i \, {\rm T} \exp\left[ \frac{1}{2} \int dt \sum_{i=1}^N \sum_{a=1}^n \chi_i^a \partial_t \chi_i^a
		\, + \, \frac{J^2}{8N^3} \int dt_1 dt_2 \sum_{a, b=1}^n \left( \sum_{i=1}^N \chi_i^a(t_1) \chi_i^b(t_2) \right)^4 \, \right] \, ,
	\end{equation}
where T is a formal time ordering operator, and $a,b$ are the replica indices.
Therefore, the action of the replica SYK model is given by
	\begin{equation}
		S=-{1\over 2}\sum_{a=1}^{n}\sum_{i=1}^N\int dt\; \chi^a_i\partial_t\chi^a_i - {J^2 \over 8N^3} \sum_{a,b=1}^{n}\int dt_1 dt_2 \left( \sum_{i=1}^N \chi_i^a(t_1) \chi_i^b(t_2) \right)^4 \, .
	\end{equation}
As it is seen after the $J_{ijkl}$-integration, the $O(N)$ symmetry $\chi_i \to O_{ij} \chi_j$ becomes visible;
with the site indices $i, j$ of the original version of the model now becoming internal $O(N)$ indices of the effective one site model.
For the $O(N)$ symmetric version of the model, one can now use the (Lagrangian version of) collective field formulation of the large $N$ limiting theory 
\cite{Jevicki:1980zg,deMelloKoch:1996mj,Das:2003vw}.
One uses the bi-local (replica) collective field
\begin{equation}
\psi^{ab}(t_1,t_2)\equiv \sum_{i=1}^N \chi^a_i(t_1)\chi^b_i(t_2)\equiv\Psi(X,Y)
\end{equation}
where a compact notation which packages (space-)time $t$ and the replica index $a$ into one variable $X$. In this notation, the collective bi-local field is anti-symmetric :
\begin{equation}
\Psi(X,Y)=-\Psi(Y,X)
\end{equation}
and its dynamics is governed by the effective action 
\begin{equation}
S_{\text{col}}= {N\over 2} \sum_X \partial_{t_1} \Psi(X, Y) \big|_{Y \to X}  + {N\over 2} \Tr (\log\Psi)-{J^2 N\over 8} \sum_{X,Y}  [\Psi(X,Y)]^4\label{eq:collective action}
\end{equation}
Here the log term summarizes the entropy of the original degrees of freedom, which are replaced by the bi-local fields.
The special usefulness of these comes from the fact that they represent a closed set under Large $N$ Schwinger-Dyson equations. 
In vector type models one can generally employ bi-local fields of various type, in particular through Gaussian Hubbard-Stratonovich type transformations, and that has been employed in the present model (see\cite {Georges:1999}). The single collective field representation provides however a compact formulation which generates the $1/N$ rules in concrete terms, with $1/N$ vertices taking a star-product form.
For this one first identifies the background field associated with variation of $S_{\text{col}}$
	\begin{equation}
		{\delta S_{\text{col}}\over \delta \Psi (X,Y)}={N\over 2}D(Y,X) + {N\over 2}[\Psi_0^{-1}](Y,X) -{J^2N\over 2} [\Psi_0(X,Y)]^3=0
	\end{equation}
Here we have defined a differential operator $D(X,Y)$ in bi-local space:
\begin{equation}
D(X,Y)\equiv \delta_{a,b}\delta(x-y)\partial_y
\end{equation}
which is seen to be anti-symmetric :
\begin{equation}
D(X,Y)=-D(Y,X)
\end{equation}
Manipulating this equation, one can obtain a recursion relation for the background field~$\Psi_0$:
	\begin{equation}
		\Psi_0(X,Y) \, = \, -D^{-1}(X,Y)-J^2\sum_{Z,W} D^{-1}(X,Z)[\Psi_0(Z,W)]^3\Psi_0(W,Y) \, .
	\label{saddle-point eq}
	\end{equation}
One can use a replica diagonal Ansatz for the background field:
	\begin{equation}
		\Psi_0(X,Y) \, = \, \delta_{a,b} \psi_0(x,y) \, .
	\end{equation}
In the strong coupling limit $J \to \infty$ required to reach the IR fixed point, one can neglect the left-hand side of the saddle point equation (\ref{saddle-point eq}) and finds
	\begin{equation}
		\psi_0(x,y) \, = \, - \left( \frac{1}{4\pi J^2} \right)^{\frac{1}{4}} \, \frac{\text{sgn}(x-y)}{\sqrt{|x-y|}} \, .
	\end{equation}
Therefore, from the replica limit (\ref{replica trick}), one obtains the tree-level free energy as
	\begin{equation}
		F_{(0)} \, = \, \frac{N}{2} \int dt \, \ln \psi_0(t,t) \, - \, \frac{NJ^2}{8} \int dt_1 dt_2 \; \psi_0^4(t_1, t_2) \, .
	\end{equation}

Expanding the collective filed around the background field
	\begin{equation}
		\Psi(X,Y) \, = \, \Psi_0(X,Y) \, + \, \sqrt{{2\over N}} \, \eta(X,Y) \, ,
	\label{background shift}
	\end{equation}
one obtains the vertices for the systematic $1/N$ expansion 
	\begin{equation}
		S \, = \, N S_{(0)} + \, S_{(2)} + \frac{1}{\sqrt{N}} \, S_{(3)} + \, \mathcal{O}(N^{-1}) \, ,
	\end{equation}
which can be expressed by a star product defined by $A\star B\equiv\int dt'A(t_1, t') B(t', t_2)$.
The quadratic term in the action reads
	\begin{eqnarray}
		&&S_{(2)}=-{1 \over 2} \Tr (\Psi_0^{-1}\star \eta\star\Psi_0^{-1}\star \eta )-{3J^2\over 2} \sum_{X,Y} \; [\Psi(X,Y)]^2[\eta(X,Y)]^2\cr
		&=&{1\over 2}\sum_{X_1,X_2,X_3,X_4} \eta(X_1,X_2) K(X_1,X_2;X_3,X_4) \eta(X_3,X_4) \, ,
	\label{eq:quadratic action}
	\end{eqnarray}
where the kernel $K(X_1,X_2;X_3,X_4)$ identified as:
	\begin{align}
		K(X_1,X_2;X_3,X_4)&=-{1\over 2}\left[\Psi_0^{-1}(X_4,X_1)\Psi_0^{-1}(X_2,X_3)-\Psi_0^{-1}(X_4,X_2)\Psi_0^{-1}(X_1,X_3)\right]\cr
		&-3J^2 \Psi_0(X_1,X_2)\Psi_0(X_3,X_4){1\over 2}\left[\delta_{X_1,X_3}\delta_{X_2,X_4}+\delta_{X_1,X_4}\delta_{X_2,X_3}\right] \, .
	\label{eq:quadratic kernel}
	\end{align}
The cubic term in the action action 
	\begin{eqnarray}
		&&S_{(3)} = \sqrt{2\over N} \left({1\over 3}\Tr (\Psi_0^{-1}\star \eta\star \Psi_0^{-1}\star \eta\star\Psi_0^{-1}\star \eta)-J^2\sum_{X,Y}\Psi_0(X,Y) [\eta(X,Y)]^3\right)\cr
		&=& {1\over 3}\sqrt{{2\over N}} \int \left[\prod_{i=1}^6 dX_i\right]\; V_3(X_1,X_2;X_3,X_4;X_5,X_6) \eta(X_1,X_2)\eta(X_3,X_4)\eta(X_5,X_6) \, ,
	\label{cubic action}
	\end{eqnarray}
specifies the cubic vertex $V_3(X_1,X_2;X_3,X_4;X_5,X_6)$ , given by
	\begin{align}
		&V_3(X_1,X_2;X_3,X_4;X_5,X_6)\cr
		=&{1\over 2^3 (3!)}\left[\Psi_0^{-1}(X_6,X_1)\Psi_0^{-1}(X_2,X_3)\Psi_0^{-1}(X_4,X_5)+(\mbox{sym})\right]\cr
		&-{3J^2\over 2^3 (3!)}\left[\Psi_0(X_1,X_2)\delta(X_3,X_5)\delta(X_4,X_6)\delta(X_5,X_1)\delta(X_6,X_2)+(\mbox{sym})\right] \, .
	\end{align}
Here, we need anti-symmetrization in $X_i$ and $X_{i+1}$ ($i=1,3,5$), and symmetrization among $(X_1,X_2)$, $(X_3,X_4)$ and $(X_5,X_6)$.
This will simplify the calculation because the bi-local fluctuation is anti-symmetric.

\subsection{1/N Diagrams}
\label{app: 1/N diagrams}
Expressing the quadratic action in terms of (space-)time and replica index explicitly, one finds that the potential term contains only diagonal piece in replica space.
Therefore, one can separate the quadratic action into the diagonal and off-diagonal parts as
	\begin{align}
		S_{(2)} \, &= \, - \, \frac{1}{2} \int \prod_{i=1}^4 dt_i \, \sum_{a=1}^n \, \eta^{aa}(t_1, t_2) \, \mathcal{K}(t_1, t_2; t_3, t_4) \, \eta^{aa}(t_3, t_4) \nonumber\\
		&\hspace{30pt} - \, \int \prod_{i=1}^4 dt_i \, \sum_{a>b} \, \eta^{ab}(t_1, t_2) \, \mathcal{K}'(t_1, t_2; t_3, t_4) \, \eta^{ba}(t_3, t_4) \, ,
	\end{align}
where
	\begin{align}
		\mathcal{K}(t_1, t_2; t_3, t_4) \, &= \, \frac{1}{2} \Big[ \psi_0^{-1}(t_4, t_1) \psi_0^{-1}(t_2, t_3) - \psi_0^{-1}(t_4, t_2) \psi_0^{-1}(t_1, t_3) \Big] \nonumber\\
		&\qquad \quad+ \, 3J^2 \psi_0(t_1, t_2) \psi_0(t_3, t_4) \epsilon(t_1,t_2;t_3,t_4) \, , \nonumber\\
		\mathcal{K}'(t_1, t_2; t_3, t_4) \, &= \, \psi_0^{-1}(t_4, t_1) \psi_0^{-1}(t_2, t_3) \, ,
	\end{align}
with
	\begin{equation}
		\epsilon(t_1,t_2;t_3,t_4)\equiv {1\over 2} \left(\delta(t_1-t_3)\delta(t_2-t_4)+\delta(t_1-t_4)\delta(t_2-t_3)\right) \, .
	\end{equation}
Accordingly, we can define diagonal and off-diagonal bi-local propagators: $\mathcal{D}(t_1, t_2; t_3, t_4)$ and $\mathcal{D}'(t_1, t_2; t_3, t_4)$ respectively.
Formally these propagators are defined by
	\begin{align}
		\mathcal{D}(t_1, t_2; t_3, t_4) \, &= \, \mathcal{K}^{-1}(t_1, t_2; t_3, t_4) \, , \nonumber\\
		\mathcal{D}'(t_1, t_2; t_3, t_4) \, &= \, \mathcal{K}'^{-1}(t_1, t_2; t_3, t_4) \, ,
	\end{align}
and their explicit expressions will be given in the next section.
Similarly, the cubic action is also decomposed as
	\begin{align}
		S_{(3)} \, &= \, \sqrt{2\over N} \int \prod_{i=1}^6 dt_i \ \mathcal{V}_{(3)} \sum_{a=1}^n \, \eta_{aa}(t_1, t_2) \eta_{aa}(t_3, t_4) \eta_{aa}(t_5, t_6) \nonumber\\
		&\ + \, 2\sqrt{2\over N} \int \prod_{i=1}^6 dt_i \ \mathcal{V}_{(3)}' \sum_{a>b} \, \Big[ \eta_{aa}(t_1, t_2) \eta_{ab}(t_3, t_4) \eta_{ba}(t_5, t_6) \nonumber\\
		&\hspace{130pt} + \, \eta_{ba}(t_1, t_2) \eta_{aa}(t_3, t_4) \eta_{ab}(t_5, t_6) \, + \, \eta_{ab}(t_1, t_2) \eta_{ba}(t_3, t_4) \eta_{aa}(t_5, t_6) \Big] \nonumber\\
		&\ + \, 3! \sqrt{2\over N} \int \prod_{i=1}^6 dt_i \ \mathcal{V}_{(3)}''
		\sum_{a>b>c} \, \Big[ \eta_{ab}(t_1, t_2) \eta_{bc}(t_3, t_4) \eta_{ca}(t_5, t_6) \, + \, (\text{permutations\ of\ } (a,b,c))\Big] \, ,
	\end{align}
where 
	\begin{align}		
		\mathcal{V}_{(3)} \, &= \, \frac{1}{2^3(3!)} \, \Big[ \psi_0^{-1}(t_6, t_1) \, \psi_0^{-1}(t_2, t_3) \, \psi_0^{-1}(t_4, t_5) \, + \, (\text{sym}) \Big] \nonumber\\
		&\quad - \, \frac{3J^2}{2^3 3!} \Big[ \delta(t_{1,3}) \delta(t_{1,5}) \delta(t_{2,4}) \delta(t_{2,6}) \psi_0(t_1, t_2) \, + \, (\text{sym}) \Big]\, , \nonumber\\
		\mathcal{V}_{(3)}' \, &= \,\mathcal{V}_{(3)}'' \, = \, \frac{1}{2^3(3!)} \, \Big[ \psi_0^{-1}(t_6, t_1) \, \psi_0^{-1}(t_2, t_3) \, \psi_0^{-1}(t_4, t_5) \, + \, (\text{sym}) \Big] \, .
	\end{align}

\begin{figure}[t!]
\vspace{20pt}
	\begin{center}
		\scalebox{0.8}{\includegraphics{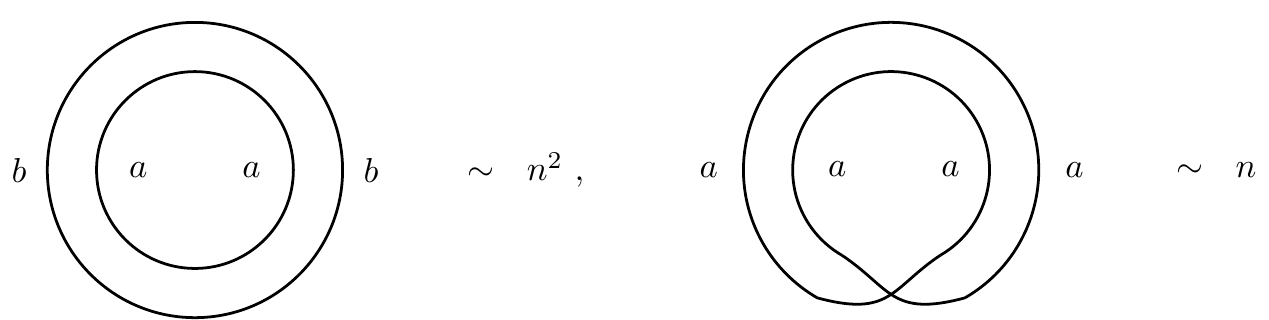}}
	\end{center}
	\vspace{-10pt}
	\caption{One-loop vacuum diagrams and their $n$-dependence.}
	\label{fig:one-loop diagrams}
\end{figure}


However, since in the replica trick (\ref{replica trick}) one has to take $n\to 0$ limit, we do not need to keep track of all contributions in the replica collective theory.
In standard {\it local} quantum field theory, this limit leads to the linked cluster theorem, i.e. one needs to consider only connected diagrams.
In the SYK model, this means that one needs only connected diagrams of the fundamental fermions $\chi_i$, as discussed in \cite{Kitaev:2015, Polchinski:2016xgd}.
In our bi-local collective theory, the $n\to 0$ limit implies that only the replica diagonal propagator is relevant after this limit.
This can be seen from explicit analysis of the diagrams, for example in Figure~\ref{fig:one-loop diagrams}.
Two-loop diagrams are given in Appendix~\ref{app:higher loop diagrams}.
There, the ``twisting'' of inside and outside lines represents diagonalization of the replica indices, i.e. $\delta_{a,b}$.
In order for a diagram to have order $n$ contribution, every loop in the diagram must have at least one twisting.
Accordingly, every bi-local propagator has the same replica index for both lines.
This concludes that for $\mathcal{O}(n)$, which survives under the replica trick, only replica diagonal propagator is relevant.
For example, the formal expression for the one-loop free energy is given by 
	\begin{equation}
		F_{(1)} \, = \, \frac{1}{2} \, \text{Tr} \log \mathcal{K} 
	\end{equation}
where the trace is taken over the bi-local time.

\subsection{Two-point Function}
\label{sec:two-point function}

The collective action based on the bi-local is known to reproduce systematically the Large $N$ results of the theory.
In particular it can be shown to generate the full set of Large $N$ Schwinger-Dyson equations \cite{Jevicki:1980zg,deMelloKoch:1996mj}.
We will now discuss in detail the case of two-point bi-local correlation function, since it plays a special role as a propagator and also for the purpose of comparison with earlier results.

The two-point function of diagonal bi-local fluctuation 
\begin{equation}
\widetilde{A}_2^{ab}(t_1,t_2;t_3,t_4)\equiv \int \left[\mathcal{D}\eta\right]\; \eta^{aa}(t_1,t_2) \eta^{bb}(t_3,t_4) e^{-S^{(2)}}
\end{equation}
satisfy the following Green's equation:
\begin{align}
&\int dt_5 dt_6\; \left[-{1\over 2}\left[\psi_0^{-1}(t_6,t_1)\psi_0^{-1}(t_2,t_5)-\psi_0^{-1}(t_6,t_2)\psi_0^{-1}(t_1,t_5)\right]\right.\cr
&\quad\left.-3J^2 \psi_0(t_1,t_2)\psi_0(t_5,t_6){1\over 2}\left[\delta(t_1-t_5)\delta(t_2-t_6)+\delta(t_1-t_6)\delta(t_2-t_5)\right]\right]
A_2^{ab}(t_5,t_6;t_3,t_4)\cr
=&{\delta_{a,b}\over 2}\left[\delta(t_1-t_3)\delta(t_2-t_4)-\delta(t_1-t_4)\delta(t_2-t_3)\right]\label{eq:green's equation for two-point function}
\end{align}
To compare with the result in \cite{Polchinski:2016xgd}, we subtract a disconnected propagator\footnote{from the point of view of the fermion $\chi_i^a$} from the two-point function of bi-local fluctuation.
\begin{equation}
\widetilde{A}_2^{ab}(t_5,t_6;t_3,t_4)\equiv {\delta_{a,b}\over 2}\left[ \psi_0(t_5,t_4)\psi_0(t_6,t_3)- \psi_0(t_5,t_3)\psi_0(t_6,t_4)\right]+\widetilde{\Gamma}^{ab}_2(t_5,t_6;t_3,t_4)
\end{equation}
We then get a differential equation for the connected propagator
\footnote{In the point of view of the fermion, $\widetilde{\Gamma}^{ab}_2$ is the connected 4-point function.} $\widetilde{\Gamma}^{ab}_2$:
\begin{align}
&\int dt_5 dt_6\; \left[-{1\over 2}\left[\psi_0^{-1}(t_6,t_1)\psi_0^{-1}(t_2,t_5)-\psi_0^{-1}(t_6,t_2)\psi_0^{-1}(t_1,t_5)\right]\right.\cr
&\quad\left.-3J^2 \psi_0(t_1,t_2)\psi_0(t_5,t_6){1\over 2}\left[\delta(t_1-t_5)\delta(t_2-t_6)+\delta(t_1-t_6)\delta(t_2-t_5)\right]\right]
\widetilde{\Gamma}_2^{ab}(t_5,t_6;t_3,t_4)\cr
=&\delta_{a,b} 3J^2 [\psi_0(t_1,t_2)]^2{1\over 2}\left[ \psi_0(t_1,t_4)\psi_0(t_2,t_3)-\psi_0(t_1,t_3)\psi_0(t_2,t_4) \right]
\end{align}
The two-point function $\widetilde{\Gamma}^{ab}_2$ represents the 4-point function of the fermion $\chi_i^a$'s (See Figure~\ref{fig:two-point function}).
Amputating the external legs of $\widetilde{\Gamma}^{ab}_2$
\begin{equation}
\widetilde{\Gamma}_2^{ab}(t_1,t_2;t_3,t_4)\equiv \int du_1du_2du_3du_4 \;\psi_0(t_1,u_1)\psi_0(t_2,u_2)\psi_0(t_3,u_3)\psi_0(t_4,u_4)\Gamma_2^{ab}(u_1,u_2;u_3,u_4),
\end{equation}
we can see a correspondence with the 1PI four -point function considered in ~\cite{Polchinski:2016xgd}
%
%
\begin{align}
&\Gamma_2^{ab}(t_1,t_2;t_3,t_4)\cr
=&\delta_{a,b} 3J^2[\psi_0(t_1,t_2)]^2{1\over 2}\left[\delta(t_1-t_3)\delta(t_2-t_4)-\delta(t_1-t_4)\delta(t_2-t_3)\right]\cr
&-3J^2 \int du_1 du_2\; [\psi_0(t_1,t_2)]^2 \psi_0(t_1,u_1)\psi_0(t_2,u_2)\Gamma_2^{ab}(u_1,u_2;t_3,t_4)
\end{align}


\begin{figure}[t!]
 \[
   \begin{minipage}[h]{0.3\linewidth}
\begin{tikzpicture}[scale=1]

\draw[thick,double] (-2,1) -- (2,1);
\draw[thick,double] (-2,-1) -- (2,-1);
\draw [fill=gray, thick, gray] (-1,1) rectangle (1,-1);
\node [above] at (-2,1) {$a$};
\node [below] at (-2,-1) {$a$};
\node [above] at (2,1) {$a$};
\node [below] at (2,-1) {$a$};
\node [below] at (-2,1) {$t_1$};
\node [above] at (-2,-1) {$t_2$};
\node [below] at (2,1) {$t_3$};
\node [above] at (2,-1) {$t_4$};

\end{tikzpicture}
\end{minipage}\;\;=\;\;\begin{minipage}[h]{0.3\linewidth}
\begin{tikzpicture}[scale=1]

\draw[thick,double] (-2,1) -- (2,1);
\draw[thick,double] (-2,-1) -- (2,-1);
\draw[thick,double] (-1,1) .. controls (-5/8,0) and (-5/8,0) .. (-1,-1);
\draw[thick,double] (-1,1) .. controls (-11/8,0) and (-11/8,0) .. (-1,-1);
\node [above] at (-2,1) {$a$};
\node [below] at (-2,-1) {$a$};
\node [above] at (2,1) {$a$};
\node [below] at (2,-1) {$a$};
\node [below] at (-2,1) {$t_1$};
\node [above] at (-2,-1) {$t_2$};
\node [below] at (2,1) {$t_3$};
\node [above] at (2,-1) {$t_4$};

\draw [fill] (0,0) circle [radius=0.05];
\draw [fill] (-0.3,0) circle [radius=0.05];
\draw [fill] (0.3,0) circle [radius=0.05];

\draw[thick,double] (1,1) .. controls (5/8,0) and (5/8,0) .. (1,-1);
\draw[thick,double] (1,1) .. controls (11/8,0) and (11/8,0) .. (1,-1);

\end{tikzpicture}
\end{minipage}\;\;-\;\;\begin{minipage}[h]{0.3\linewidth}
\begin{tikzpicture}[scale=1]

\draw[thick,double] (-2,1) -- (1.1,1);
\draw[thick,double] (-2,-1) -- (1.1,-1);
\draw[thick,double,red] (1.1,1) .. controls (2,3/4) and (1,-3/4) .. (2,-1);
\draw[thick,double,blue] (1.1,-1) .. controls (2,-3/4) and (1,3/4) .. (2,1);
\draw[thick,double] (-1,1) .. controls (-5/8,0) and (-5/8,0) .. (-1,-1);
\draw[thick,double] (-1,1) .. controls (-11/8,0) and (-11/8,0) .. (-1,-1);
\node [above] at (-2,1) {$a$};
\node [below] at (-2,-1) {$a$};
\node [above] at (2,1) {$a$};
\node [below] at (2,-1) {$a$};
\node [below] at (-2,1) {$t_1$};
\node [above] at (-2,-1) {$t_2$};
\node [below] at (2,1) {$t_3$};
\node [above] at (2,-1) {$t_4$};

\draw [fill] (0,0) circle [radius=0.05];
\draw [fill] (-0.3,0) circle [radius=0.05];
\draw [fill] (0.3,0) circle [radius=0.05];

\draw[thick,double] (1,1) .. controls (5/8,0) and (5/8,0) .. (1,-1);
\draw[thick,double] (1,1) .. controls (11/8,0) and (11/8,0) .. (1,-1);

\end{tikzpicture}
\end{minipage}
  \]
  \caption{Two-point function of (diagonal) bi-local fluctuation. The double line represents the full propagator of SYK model.}\label{fig:two-point function}
\end{figure}
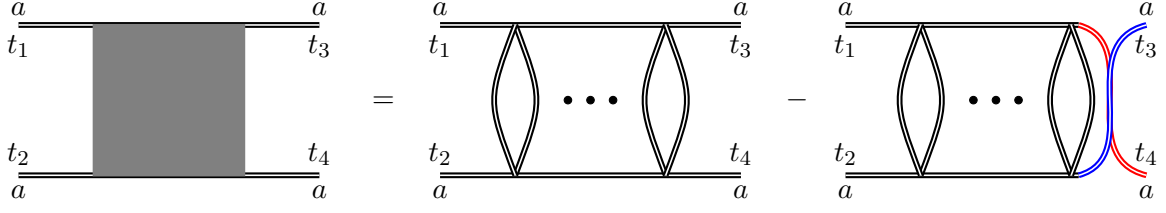


\begin{figure}[t!]
  \[
   \begin{minipage}[h]{0.3\linewidth}
\begin{tikzpicture}[scale=1.2]

        \draw[fill=gray,gray] (0:1) \foreach \x in {60,120,...,359} {
                -- (\x:1)
            }-- cycle (90:1)  ;
	\draw[thick,double] (60:1)  -- +(90:1) node[right]{$a$}node[left]{$t_2$} ;
	\draw[thick,double] (120:1) -- +(90:1) node[left]{$a$}node[right]{$t_1$};
	\draw[thick,double] (180:1) -- +(210:1) node[above]{$a$}node[below right]{$t_6$};
	\draw[thick,double] (240:1) -- +(210:1) node[below right]{$a$}node[above]{$t_5$};
	\draw[thick,double] (300:1) -- +(330:1) node[below left]{$a$}node[above]{$t_4$};
	\draw[thick,double] (0:1) -- +(330:1) node[above]{$a$}node[below left]{$t_3$};
	
	\fill (60:1)  circle[radius=2pt];
	\fill (120:1)  circle[radius=2pt];
	\fill (180:1)  circle[radius=2pt];
	\fill (240:1)  circle[radius=2pt];
	\fill (300:1)  circle[radius=2pt];
	\fill (0:1)  circle[radius=2pt];

\end{tikzpicture}
\end{minipage}\;\;=\;\;\begin{minipage}[h]{0.25\linewidth}
\begin{tikzpicture}[scale=0.8]

\draw[fill=gray,gray,shift={(240:1)},rotate=120] (0,0) rectangle (1,1);
\draw[fill=gray,gray,shift={(0:1)},rotate=240] (0,0) rectangle (1,1);

\draw[thick,double] (1,0)  .. controls +(150:1/2) and (1/2,2/3) .. (1/2,5/3);
\draw[thick,double] (-1,0)  .. controls +(30:1/2) and (-1/2,2/3) .. (-1/2,5/3);
\draw[thick,double] (240:1) -- (300:1);

\draw[thick,double,shift={(210:1)}] (180:1) -- +(210:2/3);
\draw[thick,double,shift={(210:1)}] (240:1) -- +(210:2/3);

\draw[thick,double,shift={(330:1)}] (300:1) -- +(330:2/3);
\draw[thick,double,shift={(330:1)}] (0:1) -- +(330:2/3);

\fill (180:1)  circle[radius=2pt];
\fill (240:1)  circle[radius=2pt];
\fill (180:1) +(210:1)  circle[radius=2pt];
\fill (240:1)+(210:1)  circle[radius=2pt];

\fill (300:1)  circle[radius=2pt];
\fill (0:1)  circle[radius=2pt];
\fill (300:1) +(330:1)  circle[radius=2pt];
\fill (0:1)+(330:1)  circle[radius=2pt];

\end{tikzpicture}
\end{minipage} 
\;\;+\;\;\begin{minipage}[h]{0.25\linewidth}
\begin{tikzpicture}[scale=0.8]

\draw[fill=gray,gray,shift={(120:1)}] (0,0) rectangle (1,1);
\draw[fill=gray,gray,shift={(240:1)},rotate=120] (0,0) rectangle (1,1);
\draw[fill=gray,gray,shift={(0:1)},rotate=240] (0,0) rectangle (1,1);

\draw[thick,double] (0:1) -- (60:1);
\draw[thick,double] (120:1) -- (180:1);
\draw[thick,double] (240:1) -- (300:1);

\draw[thick,double,shift={(0,1)}] (60:1) -- +(0,2/3);
\draw[thick,double,shift={(0,1)}] (120:1) -- +(0,2/3);

\draw[thick,double,shift={(210:1)}] (180:1) -- +(210:2/3);
\draw[thick,double,shift={(210:1)}] (240:1) -- +(210:2/3);

\draw[thick,double,shift={(330:1)}] (300:1) -- +(330:2/3);
\draw[thick,double,shift={(330:1)}] (0:1) -- +(330:2/3);

\fill (60:1)  circle[radius=2pt];
\fill (120:1)  circle[radius=2pt];
\fill (60:1)+(0,1)  circle[radius=2pt];
\fill (120:1)+(0,1)  circle[radius=2pt];

\fill (180:1)  circle[radius=2pt];
\fill (240:1)  circle[radius=2pt];
\fill (180:1) +(210:1)  circle[radius=2pt];
\fill (240:1)+(210:1)  circle[radius=2pt];

\fill (300:1)  circle[radius=2pt];
\fill (0:1)  circle[radius=2pt];
\fill (300:1) +(330:1)  circle[radius=2pt];
\fill (0:1)+(330:1)  circle[radius=2pt];

\end{tikzpicture}
\end{minipage}
  \]\\
  \[
  +J^2\left[-\begin{minipage}[h]{0.2\linewidth}
  \begin{tikzpicture}[scale=0.6]

\draw[thick,double,blue] (1/2,1.8) -- (1/2,0.4);
\draw[thick,double,red] (-1/2,1.8) -- (-1/2,0.4);

\draw[thick,double,red] (-1/2,0.4)  .. controls (-1/2+0.15,0) and (300:1) .. ({cos(30)+1},{-sin(30)});
\draw[thick,double,red] (-1/2,0.4)  .. controls (-1/2,0) and (240:1) .. ({cos(210)+cos(240)},{sin(210)+sin(240)});

\draw[thick,double,blue] (1/2,0.4)  .. controls (1/2,0) and (300:1) .. ({cos(30)+cos(300)},{-sin(30)+sin(300)});
\draw[thick,double,blue] (1/2,0.4)  .. controls (1/2-0.16,0) and (300:1) .. ({cos(210)+cos(180)},{sin(210)+sin(180)});

\draw[thick,double] (-1/2,0.4) -- (1/2,0.4);

\fill[blue] (1/2,0.4)  circle[radius=2pt];
\fill[red] (-1/2,0.4)  circle[radius=2pt];

\end{tikzpicture}
  \end{minipage}+
  \begin{minipage}[h]{0.2\linewidth}
\begin{tikzpicture}[scale=0.6]

\draw[fill=gray,gray,shift={(120:1)}] (0,0) rectangle (1,1);

\draw[thick,double,blue] (60:1) -- (1/2,0.4);
\draw[thick,double,red] (120:1) -- (-1/2,0.4);

\draw[thick,double,red] (-1/2,0.4)  .. controls (-1/2+0.15,0) and (300:1) .. ({cos(30)+1},{-sin(30)});
\draw[thick,double,red] (-1/2,0.4)  .. controls (-1/2,0) and (240:1) .. ({cos(210)+cos(240)},{sin(210)+sin(240)});

\draw[thick,double,blue] (1/2,0.4)  .. controls (1/2,0) and (300:1) .. ({cos(30)+cos(300)},{-sin(30)+sin(300)});
\draw[thick,double,blue] (1/2,0.4)  .. controls (1/2-0.16,0) and (300:1) .. ({cos(210)+cos(180)},{sin(210)+sin(180)});

\draw[thick,double,shift={(0,1)}] (60:1) -- +(0,2/3);
\draw[thick,double,shift={(0,1)}] (120:1) -- +(0,2/3);

\draw[thick,double] (-1/2,0.4) -- (1/2,0.4);

\fill (60:1)  circle[radius=2pt];
\fill (120:1)  circle[radius=2pt];
\fill (60:1)+(0,1)  circle[radius=2pt];
\fill (120:1)+(0,1)  circle[radius=2pt];

\fill[blue] (1/2,0.4)  circle[radius=2pt];
\fill[red] (-1/2,0.4)  circle[radius=2pt];

\end{tikzpicture}
  \end{minipage}-\begin{minipage}[h]{0.2\linewidth}
  \begin{tikzpicture}[scale=0.6]

\draw[fill=gray,gray,shift={(240:1)},rotate=120] (0,0) rectangle (1,1);
\draw[fill=gray,gray,shift={(0:1)},rotate=240] (0,0) rectangle (1,1);

\draw[thick,double,shift={(210:1)}] (180:1) -- +(210:2/3);
\draw[thick,double,shift={(210:1)}] (240:1) -- +(210:2/3);

\draw[thick,double,shift={(330:1)}] (300:1) -- +(330:2/3);
\draw[thick,double,shift={(330:1)}] (0:1) -- +(330:2/3);

\draw[thick,double,red] (-1/2,0.8)  .. controls (0,0.4) and (0.1,0.3) .. (0:1);
\draw[thick,double,red] (-1/2,0.8)  -- (240:1);

\draw[thick,double,blue] (1/2,0.8)  -- (300:1);
\draw[thick,double,blue] (1/2,0.8)  .. controls (0,0.4) and (-0.1,0.3) .. (180:1);

\draw[thick,double,red] (-1/2,0.8) -- (-1/2,1.8);
\draw[thick,double,blue] (1/2,0.8) -- (1/2,1.8);

\draw[thick,double] (-1/2,0.8) -- (1/2,0.8);

\fill (180:1)  circle[radius=2pt];
\fill (240:1)  circle[radius=2pt];
\fill (180:1) +(210:1)  circle[radius=2pt];
\fill (240:1)+(210:1)  circle[radius=2pt];

\fill (300:1)  circle[radius=2pt];
\fill (0:1)  circle[radius=2pt];
\fill (300:1) +(330:1)  circle[radius=2pt];
\fill (0:1)+(330:1)  circle[radius=2pt];

\fill[blue] (1/2,0.8)  circle[radius=2pt];
\fill[red] (-1/2,0.8)  circle[radius=2pt];

\end{tikzpicture}
  \end{minipage}+\begin{minipage}[h]{0.2\linewidth}
  \begin{tikzpicture}[scale=0.6]
  
	\draw[fill=gray,gray,shift={(120:1)}] (0,0) rectangle (1,1);
	\draw[fill=gray,gray,shift={(240:1)},rotate=120] (0,0) rectangle (1,1);
	\draw[fill=gray,gray,shift={(0:1)},rotate=240] (0,0) rectangle (1,1);
  
	\draw[thick,double,shift={(0,1)}] (60:1) -- +(0,2/3);
	\draw[thick,double,shift={(0,1)}] (120:1) -- +(0,2/3);

	\draw[thick,double,shift={(210:1)}] (180:1) -- +(210:2/3);
	\draw[thick,double,shift={(210:1)}] (240:1) -- +(210:2/3);

	\draw[thick,double,shift={(330:1)}] (300:1) -- +(330:2/3);
	\draw[thick,double,shift={(330:1)}] (0:1) -- +(330:2/3);

	\draw[thick,double,blue] (60:1) -- (1/2,0.4);
	\draw[thick,double,red] (120:1) -- (-1/2,0.4);

	\draw[thick,double,red] (-1/2,0.4)  .. controls (-1/4,0.1) and (240:0.4) .. (0:1);
	\draw[thick,double,red] (-1/2,0.4)  -- (240:1);

	\draw[thick,double,blue] (1/2,0.4)  -- (300:1);
	\draw[thick,double,blue] (1/2,0.4)  .. controls (1/4,0.1) and (300:0.4) .. (180:1);

	\draw[thick,double] (-1/2,0.4) -- (1/2,0.4);

	\fill (60:1)  circle[radius=2pt];
	\fill (120:1)  circle[radius=2pt];
	\fill (60:1)+(0,1)  circle[radius=2pt];
	\fill (120:1)+(0,1)  circle[radius=2pt];
  
	\fill (180:1)  circle[radius=2pt];
	\fill (240:1)  circle[radius=2pt];
	\fill (180:1) +(210:1)  circle[radius=2pt];
	\fill (240:1)+(210:1)  circle[radius=2pt];

	\fill (300:1)  circle[radius=2pt];
	\fill (0:1)  circle[radius=2pt];
	\fill (300:1) +(330:1)  circle[radius=2pt];
	\fill (0:1)+(330:1)  circle[radius=2pt];

	\fill[blue] (1/2,0.4)  circle[radius=2pt];
	\fill[red] (-1/2,0.4)  circle[radius=2pt];

\end{tikzpicture}
\end{minipage}   \right]
  \]
  \caption{Three-point function of (diagonal) bi-local fluctuation. Different colors (red, blue, black) of double lines are used for non-planar diagrams. They intersect only at the four-point vertices.}\label{fig:three-point function}
\end{figure}
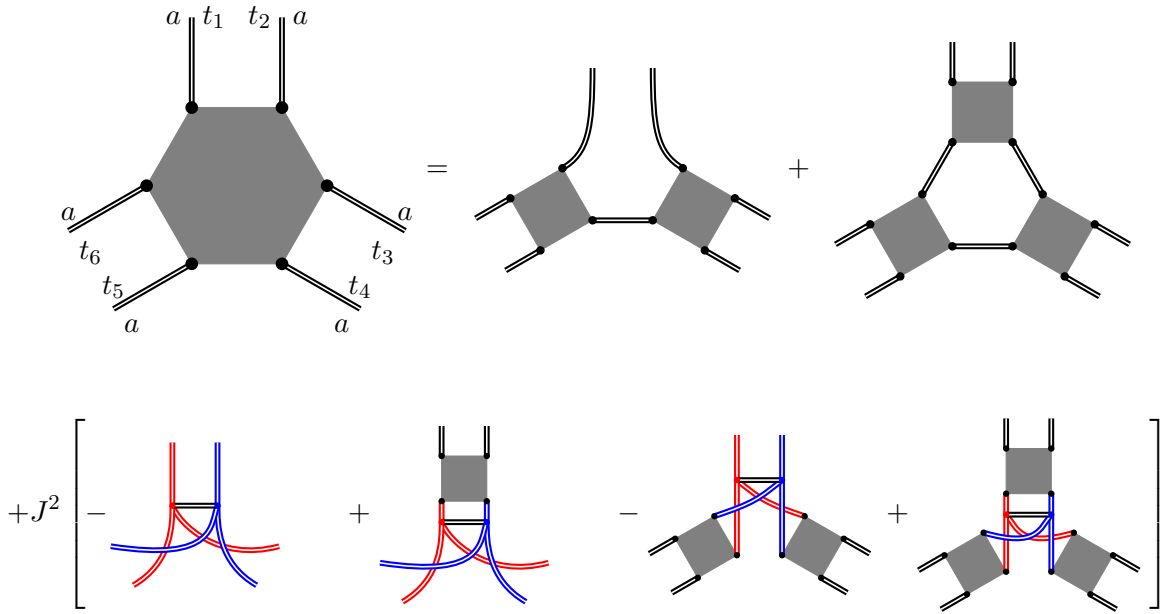

\newpage

In the Appendix~\ref{app: two-point function}, we repeat the similar calculation for the three-point function of the bi-local fluctuations, and we pick up the connected propagator in the point of view of the fermion to get $\widetilde{\Gamma}_3$. The three-point function $\widetilde{\Gamma}_3$ can be compared to the 6-point function of SYK model (See Figure~\ref{fig:three-point function}).

\section{AdS Bulk}
\label{sec:ads bulk}

From the replica collective field theory in the $n \to 0$ limit, we have the resulting theory to be given by the action (\ref{eq:collective action}).
The partition function involving the bi-local field $\Psi(t_1, t_2)$ is 
	\begin{align}
		Z \, = \, \int \prod_{t_1, t_2} \mathcal{D}\Psi(t_1, t_2) \ \mu(\Psi) \, e^{-S_{\rm col}[\Psi]} \, , 
	\label{eq:collective partition function}
	\end{align}
where the precise measure $\mu(\Psi)$ is given for example in \cite{Jevicki:2014mfa}.
The critical theory is represented by the action
	\begin{align}
		S_{\rm c}[\Psi] \, = \, \frac{N}{2} \int dt \, \log \Psi(t, t) \, - \, \frac{J^2N}{8} \int dt_1 dt_2 \, \Psi^4(t_1, t_2) \, ,
	\end{align}
which exhibits an (emergent) symmetry under $t \to f(t)$ reparametrization and 
	\begin{align}
		\Psi(t_1, t_2) \, \to \, \Psi_{f}(t_1, t_2) \, = \, \big|f'(t_1) f'(t_2) \big|^{\frac{1}{4}} \, \Psi(f(t_1), f(t_2)) \, .
	\end{align}
This symmetry is responsible for the appearance of zero modes in the strict IR critical theory,
a problem which can be addressed as in the quantization of systems with symmetry modes developed in \cite{Gervais:1975yg}.
We describe this procedure in the following subsection.

\subsection{Time as a Dynamical Variable}\label{sec:time as a dynamical variable}
The above symmetry mode representing time reparametrization can be elevated to a dynamical variable introduced following \cite{Gervais:1975yg} through the Faddeev-Popov method.
We insert into the partition function (\ref{eq:collective partition function}), the functional identity:
	\begin{align}
		\int \prod_{t} \mathcal{D}f(t) \ \prod_{t}\delta\left( \int u \cdot \Psi_f \right) \left| \frac{\delta \int u \cdot \Psi_f}{\delta f} \right| \, = \, 1 \, ,
	\end{align}
so that after an inverse change of the integration variable, it results in a combined representation 
	\begin{align}
		Z \, = \, \int \prod_{t} \mathcal{D}f(t) \prod_{t_1, t_2} \mathcal{D}\Psi(t_1, t_2) \ \mu(f, \Psi) \, \delta\left( \int u \cdot \Psi_f \right) \, e^{-S_{\rm col}[\Psi, f]} \ ,
		\label{eq:gauged collective partition function}
	\end{align}
with a Jacobian.

Consequently $f(t)$ is now introduced as a dynamical degree of freedom, at the same time the delta function condition projects out a state associated with the wave function $u(t_1, t_2)$.
This wave function is arbitrary (representing different gauges), it will be chosen to eliminate the zero mode of the IR.
The total action is now
	\begin{align}
		S_{\rm col}[\Psi, f] \, = \, \frac{N}{2} \int dt \Big[ \partial_t \Psi_f(t, t') \Big]_{t =t'} \, + \, S_{\rm c}[\Psi] \, .
	\end{align}
The action for the dynamical variable $f(t)$ is contained in the first term, and follows after a background shift (\ref{background shift}). This we consider in
 in Appendix~\ref{app:evaluation of the time action} within the scheme of an epsilon expansion starting from $q=2$, where the short distance limit is most directly evaluated, obtaining the Schwarzian derivative form:
	\begin{align}
		S[f] \, = \, - \, \frac{N}{24\pi J} \int dt \, \left[ \, \frac{f'''(t)}{f'(t)} \, - \, \frac{3}{2} \left( \frac{f''(t)}{f'(t)} \right)^2 \, \right] \, \equiv \, - \, \frac{N}{24\pi J} \int dt \ \{ f, t\} \, .
	\end{align}
One can then consider corrections through a $\varepsilon \equiv (q-2)/2$ expansion.
There will also be corrections induced by the correction (due to $J$) of the background field. 
All these are only expected to change overall coefficient in front of the Schwarzian derivative to $\alpha(\varepsilon)$:
	\begin{align}
		S[f] \, = \, - \, \frac{N\alpha}{24\pi J} \int dt \ \{ f, t\} \, ,
	\label{Schwarzian action}
	\end{align}
whose value can be read of from a numerical evaluation of the specific heat performed in \cite{Maldacena:2016hyu}.

For physical transparency, we can instead of $f(t)$ introduce $T(f)$ as a dynamical variable.
Namely, we change $f(t) \to T(f)$, where $T=t$.
For that, the action (\ref{Schwarzian action}) becomes
	\begin{align}
		S[T] \ = \ \frac{N\alpha}{48\pi J} \int df \ T' \, \left( \frac{T''}{(T')^2} \right)^2 \ = \ \frac{N\alpha}{48\pi J} \int d\tau \, \frac{(T'')^2}{(T')^3} \, ,
	\end{align}
Here and in the following, we use $\tau$ for the integration variable instead of $f$, and the prime denotes $\tau$ derivative.

It is useful to perform a canonical quantization of this system and construct a conserved energy as a conjugate to time.
We introduce an auxiliary variable 
	\begin{align}
		q (\tau) \, \equiv \, \frac{T''}{(T')^3} \, ,
	\label{define-q}
	\end{align}
and 
	\begin{align}
		S[T] \, &= \, \frac{N\alpha}{24\pi J} \int d\tau \left[ \, T'' \, q \, - \, \frac{1}{2} (T')^3 q^2 \, \right] \nonumber\\
		&= \, - \, \frac{N\alpha}{24\pi J} \int d\tau \left[ \, T' \, q' + \, \frac{1}{2} (T')^3 q^2 \, \right] \, \equiv \, S[T, q] \, .
	\end{align}
The conjugate variables are denoted by $\Pi_{q}$ and $\Pi_{T}$ respectively.
Then the Hamiltonian is given by 
	\begin{align}
		H \, = \, \Pi_T \, T' \, + \, \Pi_q \, q' \, - \, L \, = \, - \, \frac{24\pi J}{N \alpha} \, \left[ \, \Pi_T \Pi_q \, + \, \frac{12\pi J}{N\alpha} \, q^2 \, \Pi_q^3 \, \right] \, .
	\end{align}
This is a ``Hamiltonian'' for $\tau$-dynamics, but $\Pi_T$, which is conjugate to $T(\tau)$ is the energy:
	\begin{align}
		E \, = \, \Pi_T \, .
	\end{align}
From the Euler-Lagrange equation of $T$, one can see that the energy is conserved:
	\begin{align}
		\frac{dE}{d\tau} \, = \, 0 \, .
	\label{energy-conservation}
	\end{align}

We note that the configuration 
	\begin{align}
		T_{\beta}(\tau) \, = \, \frac{\beta}{\pi} \, \arctan(\tau) \, ,
	\label{classical-solution}
	\end{align}
(the inverse of $f(t)=\tan(\pi t/\beta)$) represents a {\it classical solution} of the $S[T]$ action.
For this classical solution, the energy is a constant
	\begin{align}
		E(T_{\beta}(\tau)) \, = \, \frac{N\alpha}{12\pi J} \, \frac{\pi^2}{\beta^2} \, .
	\end{align}

\subsection{The Propagator}
\label{sec:the propagator}

Let us now consider the quadratic action in the strong coupling limit $J |t| \gg 1$ given in \eqref{eq:quadratic action}.
\begin{align}
S^{(2)}&=-{1 \over 2} \Tr (\psi_0^{-1}\star \eta\star\psi_0^{-1}\star \eta )-{3J^2\over 2} \int dt_1 dt_2 \; [\psi_0(t_1,t_2)]^2[\eta(t_1,t_2)]^2\cr
&={1\over 2}\int dt_1dt_2dt_3dt_4\; \eta(t_1,t_2) K(t_1,t_2;t_3,t_4) \eta(t_3,t_4)
\end{align}
%
%
%
The bi-local field $\eta(t_1,t_2)$ will be identified as AdS$_2$ bulk field:
\begin{equation}
\eta(t_1,t_2)=\Phi(t,z)
\end{equation}
with the following coordinate transformation from $(t_1,t_2)$ to $(t,z)$
\begin{equation}
t={1\over 2}(t_1+t_2)\quad,\quad z={1\over 2}(t_1-t_2).\label{eq:transformation}
\end{equation}
We have a mode expansion of $\eta(t_1,t_2)$ (and therefore $\Phi(t,z)$):
\begin{equation}
\Phi(t,z)=\eta(t_1,t_2)\equiv \sum_{\nu w}\widetilde{\Phi}_{\nu w} u_{\nu w}(t_1,t_2)\label{eq:expansion of eta}
\end{equation}
in terms of complete basis $u_{\nu w}(t_1,t_2)$ defined by
\begin{equation}
u_{\nu w}(t_1,t_2)=e^{iw{t_1+t_2\over 2}}\sgn(t_1-t_2)Z_\nu\left(|w(t_1-t_2)/ 2|\right)
\end{equation}
Here, $Z_\nu(z)$ is a linear combination of two Bessel functions found in \cite{Polchinski:2016xgd} :
\begin{eqnarray}
&Z_\nu(x)\equiv J_\nu(x)+\xi_\nu J_{-\nu}(x)\qquad \mbox{where}\quad \xi_\nu= {\tan{\nu\pi\over 2}+1\over \tan{\nu\pi\over 2}-1}\\
&\left(x^2\partial_x^2 +x\partial_x -x^2\right)Z_\nu(x)=\nu^2 Z_\nu(x)
\end{eqnarray} 
Using the identity which will be proven in Appendix~\ref{app:identity}
\begin{align}
\int {dt_3dt_4\over \left|{t_3-t_4}\right|}\; \psi_0(t_1,t_3)\psi_0(t_2,t_4) u_{\nu w}(t_3,t_4)= - {16\sqrt{\pi} \over 3J} g(\nu) u_{\nu w}(t_1,t_2),
\end{align}
where $g(\nu)=-(3/2\nu) \tan(\pi \nu/2)$ defined in~\cite{Polchinski:2016xgd}, \eqref{eq:expansion of eta} can be rewritten as
\begin{equation}
 \Phi(t,z) =- \sum_{\nu w} { 3J \widetilde{g}(\nu) \over 16\sqrt{\pi}  }\widetilde{\Phi}_{\nu w} \int {dt_3dt_4\over \left|{t_3-t_4}\right|}\; \psi_0(t_1,t_3)\psi_0(t_2,t_4)u_{\nu w} (t_3,t_4) 
\end{equation}
where we define 
\begin{equation}
\widetilde{g}(\nu)\equiv {1\over g(\nu)}
\end{equation}

We will also require for our evaluation the diagonalization of $[\psi_0(t_1,t_2)]^2 \psi_0(t_1,t_3)\psi_0(t_2,t_4)$ found in \cite{Polchinski:2016xgd}, which introduced the eigenvalue $g(\nu)$.

Applying the above relation to one of $\eta$ in each term of quadratic action, one can diagonalize the quadratic action as follows.
\begin{align}
S^{(2)}=&{1\over 2} \sum_{\nu,\nu'}\int dwdw'\; {3J \widetilde{g}(\nu')\over 16\sqrt{\pi}  }\widetilde{\Phi}_{\nu w} \widetilde{\Phi}_{\nu' w'}   \int  {dt_1dt_2 \over \left|{t_1-t_2}\right|}   \; u_{\nu w}(t_1,t_2) u_{\nu' w'}(t_1,t_2) \cr
&+{3J^2\over 2} \sum_{\nu,\nu'} \int dw dw'\;  {3J \widetilde{g}(\nu') \over 16\sqrt{\pi} } \widetilde{\Phi}_{\nu w} \widetilde{\Phi}_{\nu' w'}\int {dt_1dt_2 dt_3 dt_4\over  \left|{t_3-t_4}\right|}\; \cr
&\qquad\qquad\times u_{\nu w}(t_1,t_2)[\psi_0(t_1,t_2)]^2 \psi_0(t_1,t_3)\psi_0(t_2,t_4)u_{\nu' w'}(t_3,t_4)
\end{align}
Using next the diagonalization of $[\psi_0(t_1,t_2)]^2 \psi_0(t_1,t_3)\psi_0(t_2,t_4)$ we obtain\footnote{From the complex conjugate of the basis $u_{\nu w}$ 
\begin{equation}
u^*_{\nu w}(t,z)=\xi^*_\nu u_{\nu,-w}\quad\mbox{for}\;\; \nu=ir\qquad (r>0),
\end{equation} 
the reality condition of the bi-local fluctuation $\eta(t_1,t_2)=\Phi(t,z)$ implies that
\begin{equation}
\Phi_{\nu w}^*=\xi_\nu \Phi_{\nu,-w}\quad\mbox{for}\;\; \nu=ir\qquad (r>0).
\end{equation}
}
\begin{equation}
S^{(2)}
={J\over 2} \sum_{\nu, w}  {3 N_\nu \over 16\sqrt{\pi} } \widetilde{\Phi}_{\nu, w}^*  \left(\widetilde{g}(\nu) - 1\right)\widetilde{\Phi}_{\nu, w} \label{eq:quadratic action of collective field1}
\end{equation}
where $N_\nu$ is the normalization of $u_{\nu w}(t_1,t_2)$~\cite{Polchinski:2016xgd}
\begin{equation}
N_\nu=\begin{cases}
\;\;{2\pi \over \nu}& \;\; \mbox{for}\;\;\nu={3\over 2}+2n\quad (n=0,1,\cdots)\\
\;\;{8\pi \sin \pi\nu \over \nu}&\;\;\mbox{for}\;\; \nu=ir\quad(r>0)\\
\end{cases}.
\end{equation}
This leads to a propagator
\begin{align}
\mathcal{D}(t,z;t',z')=&\int_0^{\infty} dr \int dw\; {16\sqrt{\pi}\over 3JN_{ir} } { u_{ir, w}^*(t, z) \, u_{ir, w}(t',z') \over \widetilde{g}(ir)-1}\cr
&+\sum_{\substack{\nu=3/2+2n\\ n=0,1,2,\cdots}}\int dw\; {16\sqrt{\pi}\over 3JN_\nu} { u_{\nu w}^*(t, z) \, u_{\nu w}(t',z') \over \widetilde{g}(\nu)-1}
\label{bi-local propagator}
\end{align}
This propagator is different in detail from the one considered in \cite{Polchinski:2016xgd},
in particular $\widetilde{g}$ replaces $g$ and that the dependence on the coupling constant $J$ is different. 
This is related to the effect of amputation of external legs for 1PI case of \cite{Polchinski:2016xgd} as we discussed in the previous section.
It is the above propagator howerer that defines the $1/N$ Feynman rules.
(Note that the poles appearing in the two propagators are identical.)
The result of the bi-local propagator (\ref{bi-local propagator}) shows a divergence at the pole given by $\nu=3/2$.
This zero mode divergence in the resulting propagator corresponds to the symmetry (or Goldstone modes \cite{Maldacena:2016, Maldacena:2016hyu})
associated with conformal reparameterization symmetry, $(t_1, t_2) \to (f(t_1), f(t_2))$ in the strong coupling limit $J|t|\gg1$.
The zero mode is eliminated by our delta function constraint in (\ref{eq:gauged collective partition function}).
\footnote{
In (\ref{eq:gauged collective partition function}), we take
$u = \int dt \, e^{i \omega t} \frac{\delta \Psi^0_f(t_1, t_2)}{\delta f(t)} \propto  u_{\frac{3}{2}, \omega}(t_1, t_2) \sim J_{\frac{3}{2}}( |\omega(t_1-t_2)|/2 )$.}

We note that the continuous sum of the bi-local propagator (first line of (\ref{bi-local propagator})) is reminiscent of the 2D non-critical string loop propagators \cite{Moore:1991ag}.
One can evaluate the integral as in \cite{Moore:1991ag} and reduce it to sum of poles.
%
%
%
%
Since $\nu= ir$ $(r\in \mathbb{R})$, one can rewrite the term as
\begin{align}
	&\ \int_{-i\infty}^{i\infty} {d\nu\over i} \int_{-\infty}^{\infty} dw\; {8\sqrt{\pi}\over 3JN_\nu} { u_{\nu w}^*(t, z) \, u_{\nu w}(t',z') \over \widetilde{g}(\nu)-1}\cr
	=&\ i \ \frac{\sgn(zz')}{3 \sqrt{\pi} J} \int_{-i\infty}^{i\infty} d\nu \int_{-\infty}^{\infty} dw\ \nu \, e^{-iw(t-t')} \,
	\frac{\big[J_{-\nu}(|wz|) + \xi_{-\nu} J_{\nu}(|wz|)\big] J_{\nu}(|wz'|)}{\cos {\pi \nu\over 2}\left[{2\nu \over 3}\cos {\pi \nu \over 2}+\sin{\pi \nu \over 2}\right]} \, .\label{eq:u into bessel function}
\end{align}
where we assume $z>z'$ without loss of generality.\footnote{Note that the Bessel functions in \eqref{eq:u into bessel function} behave well in the limit $z\longrightarrow \infty$ and $z'\longrightarrow 0$: 
\begin{alignat}{4}
J_\nu(z)\sim& z^\nu&&\mbox{for} \quad z\ll1\\
J_{-\nu}(z)+\xi_{-\nu}J_\nu (z) \sim & {z^{-{1\over 2} }\cos z \sin {\pi \nu \over 2}\over 1+\tan{\pi\nu\over 2} } &\qquad &\mbox{for}\quad z\gg 1
\end{alignat}}
In order to evaluate the $\nu$-integral as a contour integral in the complex $\nu$ plane, we consider poles of the integrand on this complex plane.
Even though $\cos(\pi \nu/2)$ in the denominator becomes zero at $\nu=2n+1$ $(n\in \mathbb{Z})$, these points are not poles because $(-1)^{\nu}+\xi_{-\nu} =0$ at these points.
%
%
%
On the other hand, $\xi_{-\nu}$ has simple poles at $\nu=2n+{3\over 2}$.
The other set of poles of the integrand is given by the solutions, $p_m$ of the following equation:
\begin{equation}
	{2p_m \over 3} \, = \, - \tan \left(\pi p_m\over 2\right) \, , \qquad 2m+1<p_m<2m+2 \quad (m\in \mathbb{Z})
\end{equation}
If we close the contour in $\Re (\nu)\ge 0$ region, the integral picks up simple poles at $\nu=p_m$ ($m>0$) and $\nu=2n+{3\over 2}$ ($n=1,2,\cdots$).
The residue contributions from all poles in $\Re (\nu)\ge 0$ region 
is given by
	\begin{align}
		&- \, \text{sgn}(zz') \frac{8}{3\sqrt{\pi}J} \, \sum_{\substack{n=1\\ \nu=3/2+2n}}^{\infty} \int_{-\infty}^{\infty} d\omega \, \nu \, e^{-i\omega(t-t')} \,
		\frac{J_{\nu}(|\omega z|)J_{\nu }(|\omega z'|)}{\frac{2\nu}{3}-1} \nonumber\\
		&- \, \sgn(zz') \, {4\over \sqrt{\pi}J} \sum_{m=1}^\infty \int_{-\infty}^{\infty} dw\; \frac{e^{-iw(t-t')}}{\sin(\pi p_m )} \, \frac{p_m^2}{p_m^2+(3/2)^2}
		\left[J_{-p_m}(|wz|)  + {p_m+{3\over 2}\over p_m-{3\over 2}} J_{p_m}(|wz|) \right]J_{p_m}(|wz'|) \, .
	\end{align}
We can see that the first line contribution exactly cancels with the discrete sum of the bi-local propagator (\ref{bi-local propagator}). 
This leaves the sum of poles of $p_m$ $(m>0)$ as in \cite{Maldacena:2016hyu}.
Our final expression of the bi-local propagator is now given by
	\begin{align}
		\mathcal{D}(t,z;t',z') \, &= \, 
		\, - \, \sgn(zz') \, {4\over \sqrt{\pi}J} \sum_{m=1}^\infty \int_{-\infty}^{\infty} dw\; \frac{e^{-iw(t-t')}}{\sin(\pi p_m )} \, \frac{p_m^2}{p_m^2+(3/2)^2} \nonumber\\
		&\hspace{110pt} \times \left[J_{-p_m}(|wz|)  + {p_m+{3\over 2}\over p_m-{3\over 2}} J_{p_m}(|wz|) \right]J_{p_m}(|wz'|) \, .
	\label{eq:propagator final form}
	\end{align}

At each pole, we have one AdS-type contribution. For each pole, one may consider an effective action of a scalar field with mass, $M^2_m=p_m^2-{1\over 4}$, ($m>0$) in AdS$_2$:
\begin{equation}
S^{\text{eff}}_m={1\over 2} \int \sqrt{|g|} \, dx^2 \left[-g^{\mu\nu} \partial_\mu\phi_m \partial_\nu\phi_m -\left(p_m^2-{1\over 4}\right)\phi_m^2\right]\label{eq:effective action for scalar field 1}
\end{equation}
where the metric $g_{\mu\nu}$ is given by
\begin{equation}
g_{\mu\nu}=\mbox{diag}(-1/z^2,1/z^2)\ .
\end{equation}
Since that $p_m$ is a zero of $\widetilde{g}(\nu)-1$, we can define $f_m(\nu)$ by 
%
%
\begin{align}
\widetilde{g}(\nu)-1&=-{2\nu\over 3 }\cot {\nu \pi \over 2}-1
\equiv\left[\nu^2-(p_m)^2\right]f_m(\nu) \ .
\end{align}
%
%
%
%
%
%
%
%
Note that $\widetilde{g}(\nu)-1$ and $f_m(\nu)$ are even functions of $\nu$. Recalling the Bessel's differential equation
\begin{equation}
\left[z^2\partial_z^2+z\partial_z +w^2z^2\right] J_{\pm\nu}(wz)= \nu^2 J_{\pm\nu}(wz)\ ,
\end{equation}
we define $\Phi_m(t,z)$ by a field redefinition as follows.
 \begin{equation}
 \phi_m(t,z)\equiv  z^{1\over 2} \sqrt{f(\sqrt{\mathsf{D}_{\rm B}})}  \Phi_m(t,z)
 \end{equation}
where Bessel differential operator $\mathsf{D}_{\rm B}$ is given by
\begin{equation}
\mathsf{D}_{\rm B}\equiv z^2\partial_z^2+z\partial_z -z^2\partial_t^2
\end{equation}
The appearance of a non-polynomial differential operator defining the scalar field propagation is not unusual in string theory,
in the present case one expects gravitational fields which are integrated out in the above representation.
We also note a similarity with the $c=1$ Noncritical string duality case where one defines propagators associated with loop variables of the matrix model \cite{Moore:1991ag}.
In that case through external leg redefinition, one obtains a local picture with a standard Laplacian of \cite{Das:1990kaa}.
Here we can also introduce nonlocal redefinitions of the field to achieve an analogues result relating a standard form of the AdS field theory to collective field theory. By the field redefinition, the effective action can be written as
\begin{equation}
S^{\text{eff}}_m={1\over 2}  {3J \over 8\sqrt{\pi}  } \int dt \int_0^\infty {dz\over z} \;\Phi(t,z) \left[\widetilde{g}(\sqrt{\mathsf{D}_{\rm B}} )-1 \right]  \Phi(t,z)\ .\label{eq:effective action for scalar field 2}
\end{equation}
Note that this has the same form as the quadratic action of collective field in \eqref{eq:quadratic action of collective field1} after we express \eqref{eq:quadratic action of collective field1} in the coordinate space. When one takes the pole at $p_m$, the effective action $S^{\text{eff}}_m$ in \eqref{eq:effective action for scalar field 2} leads to the on-shell propagator with $\nu=p_m$ in \eqref{eq:propagator final form}.

The cubic interaction in \eqref{cubic action} consists of two terms. The first one shows bi-locality
\begin{equation}
{1\over 3} \sqrt{2\over N} \Tr (\psi_0^{-1}\star \eta\star \psi_0^{-1}\star \eta\star\psi_0^{-1}\star \eta)= \sqrt{2\over N} J^{3/2}\!\sum_{\nu_1,\nu_2,\nu_3}\! \int  \left[\prod_{i=1}^3dw_i\right] c_{\nu_1 \nu_2\nu_3}^{w_1w_2w_3} \widetilde{\Phi}_{\nu_1,w_1}\widetilde{\Phi}_{\nu_2,w_2}\widetilde{\Phi}_{\nu_3,w_3}\label{eq:bilocal cubic interaction}
\end{equation}
where $c_{\nu_1 \nu_2\nu_3}^{w_1w_2w_3}$ is defined by
\begin{equation}
c_{\nu_1 \nu_2\nu_3}^{w_1w_2w_3}=-{1\over 16\pi^{1/2} }\widetilde{g}(\nu_2) \int {\left[\prod_{i=1}^4 dt_i\right]\over |t_3-t_4|} \;
{\psi_0^{-1}(t_1,t_2)\over J^{1/2} } u_{\nu_1 w_1}(t_2,t_3) u_{\nu_2 w_2}(t_3,t_4)u_{\nu_3 w_3}(t_4,t_1)
\end{equation}
On the other hand, the second term is local in bulk:
\begin{equation}
-J^2\sqrt{2\over N} \int dt_1 dt_2\; \psi_0(t_1,t_2)\left[\Phi(t_1,t_2)\right]^3=-{(2J)^{3/2}\over \pi^{1/4}}\int dt\int_0^\infty {dz\over z^2}\; [z^{1\over 2}\Phi(t,z)]^3
\end{equation}
The same features characterize the 4-point interactions. For the higher point ($n=5,6,7,\cdots$) interactions only the bi-local trace (star product) type term appears \eqref{eq:bilocal cubic interaction}.

\section{Conclusion}
\label{sec:conclusion}

We have defined the $O(N)$ singlet sector of the SYK model in terms of bi-local field $\Psi$ and a dynamical time coordinate $f(t)$.
The latter introduces a time-reparametrization gauge symmetry which is used to define the theory in the IR limit and eliminate the associated zero mode problem.
This exact representation of the bi-local theory with a dynamical time variable and a reparametrization symmetry which is present even away from the IR exhibits similarity with AdS gravity and will be of relevance for establishing the dual correspondence \cite{Almheiri:2014cka, Jensen:2016pah, Maldacena:2016upp, Engelsoy:2016xyb}.
There are a number of topic that can be part of further studies. First, concerning the 3-vertex (and also higher point vertices) the question of AdS locality is of definite interest. Some indications that this might be possible to achieve were given, this question is also closely related to the external leg redefinitions of the scalar field. A continuing study of finite temperature extension of the formalism is of major relevance. For this, the formalism of TFD (Thermo-Field dynamics) applies and can be implemented in parallel to recent TFD $O(N)$ vector model studies \cite{Jevicki:2015sla,Jevicki:2015pza}.

\acknowledgments

We would like to thank Yingfei Gu for some corrections.
This work is supported by the Department of Energy under contract DE-SC0010010.

\appendix

\section{Higher Loop Diagrams}
\label{app:higher loop diagrams}
In this appendix, we list discuss diagrams for the replica collective theory and their $n$ dependence up to two-loop order.
Here, the ``twisting'' of the bi-local propagator represents the diagonalization of the replica space, i.e. $\delta_{a,b}$. 
Because of this twisting, all $\mathcal{O}(n)$ diagrams contain only one replica index, which implies that only the replica diagonal propagator is relevant after the replica trick.

\begin{figure}[t!]
	\begin{center}
		\scalebox{0.8}{\includegraphics{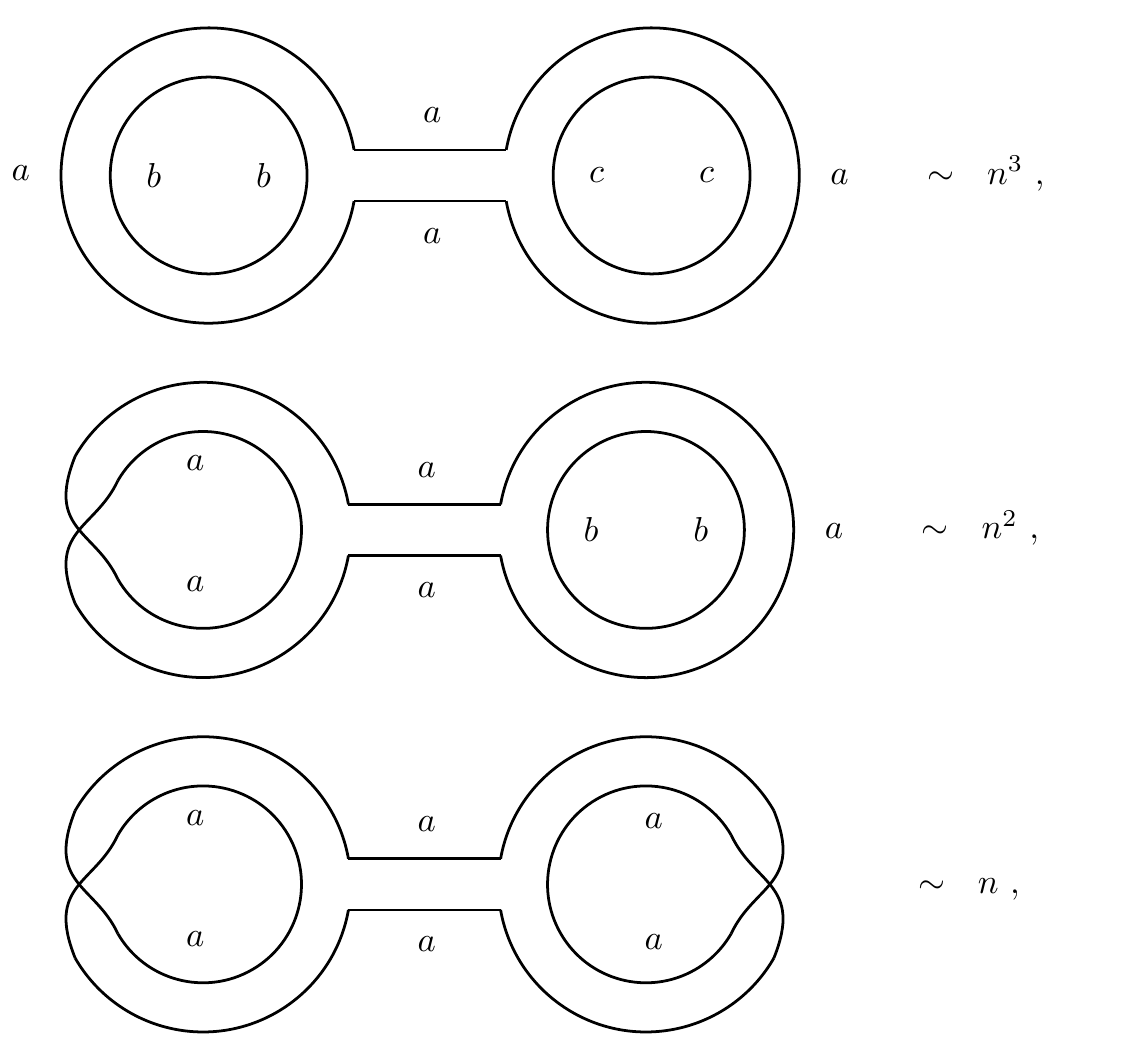}}
	\end{center}
	\vspace{-10pt}
	\caption{First kind of two-loop vacuum diagrams and their $n$-dependence.}
\end{figure}

\begin{figure}[h!]
	\begin{center}
		\scalebox{0.7}{\includegraphics{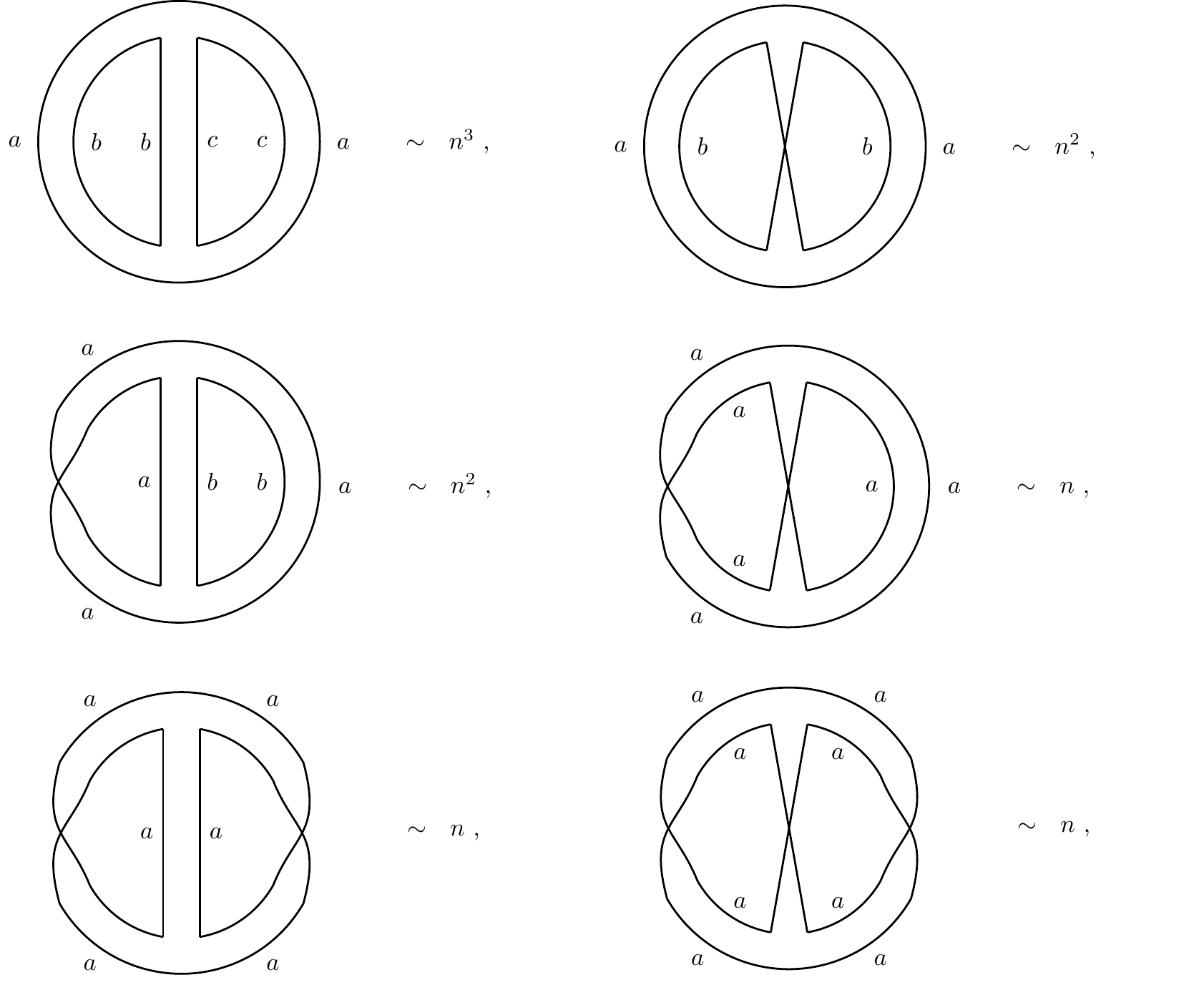}}
	\end{center}
	\vspace{-10pt}
	\caption{Second kind of two-loop vacuum diagrams and their $n$-dependence.}
	\vspace{10pt}
\end{figure}

\clearpage
\begin{figure}[t!]
	\begin{center}
		\scalebox{0.8}{\includegraphics{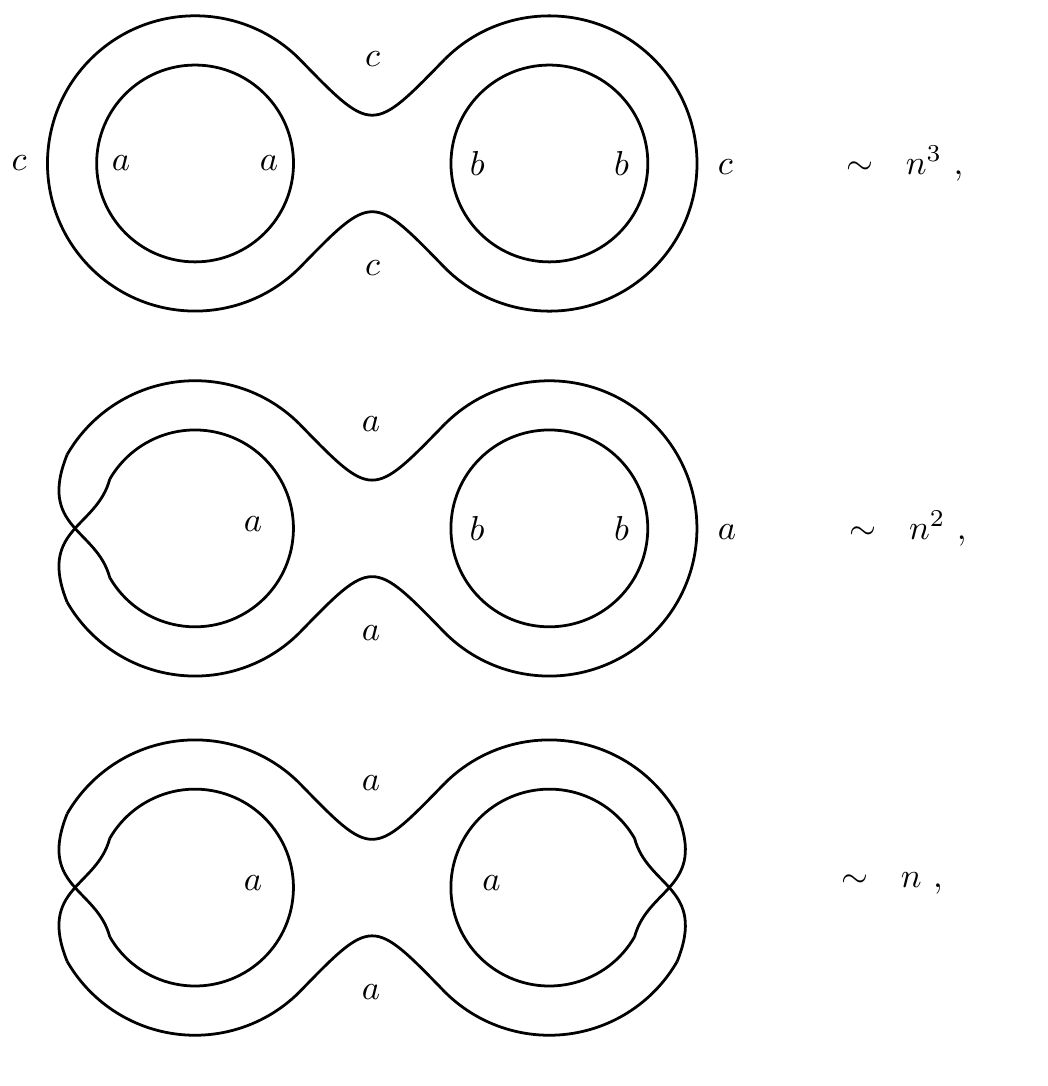}}
	\end{center}
	\vspace{-10pt}
	\caption{Third kind of two-loop vacuum diagrams and their $n$-dependence.}
\end{figure}


\section{Two- and Three-point Function}
\label{app: two-point function}

The cubic action is given by (\ref{cubic action}).
Let us consider the following identity
\begin{equation}
\int \left[\mathcal{D}\eta\right]\; {\delta\over \delta \eta(X_1,X_2)}\left(\eta(X_3,X_4)\eta(X_5,X_6)e^{-S^{(2)}-S^{(2)} }\right)=0\label{eq:three-point function identity}
\end{equation}
Defining a delta function in bi-local space
\begin{equation}
\delta(X_1,X_2;X_3,X_4)\equiv {1\over 2} \left[\delta_{X_1,X_3}\delta_{X_2,X_4}-\delta_{X_1,X_4}\delta_{X_2,X_3}\right],
\end{equation}
\eqref{eq:three-point function identity} can be written as up to $\mathcal{O}(N^{-1/2})$
\begin{eqnarray}
&&\int dU_1dU_2\; V_2(X_1,X_2;U_1,U_2) \widetilde{A}'_3(X_3,X_4;X_5,X_6;U_1,U_2)\cr
&&+\sqrt{ {2 \over N}}\delta(X_1,X_2;X_3,X_4) \int \left[\prod_{i=1}^6 dU_i\right] \; V_3(U_1,\cdots,U_6)\widetilde{A}_2(X_5,X_6;U_1,U_2)\widetilde{A}_2(U_3,U_4;U_5,U_6)\cr
&&+\sqrt{ {2 \over N}}\delta(X_1,X_2;X_5,X_6) \int \left[\prod_{i=1}^6 dU_i\right] \; V_3(U_1,\cdots,U_6)\widetilde{A}_2(X_3,X_4;U_1,U_2)\widetilde{A}_2(U_3,U_4;U_5,U_6)\cr
&&+\sqrt{2\over N} \int \left[\prod_{i=1}^4dU_i\right] \;  V_3(X_1,X_2;U_1,U_2;U_3,U_4)\cr
&&\times\left[ \widetilde{A}_2(X_3,X_4;X_5,X_6)\widetilde{A}_2(U_1,U_2;U_3,U_4)+ 2\widetilde{A}_2(X_3,X_4;U_1,U_2)\widetilde{A}_2(X_5,X_6;U_3,U_4)\right]\cr
=&&0\label{eq:equation for three point function}
\end{eqnarray}
Since we are also interested in the three-point function of diagonal bi-local fluctuations, we may use \eqref{eq:green's equation for two-point function} to solve \eqref{eq:equation for three point function} for $\widetilde{A}'_3(X_3,X_4;X_5,X_6;U_1,U_2)$.
%
%
\begin{align}
&\widetilde{A}'_3(X_1,X_2;X_3,X_4;X_5,X_6)\cr
=&-\sqrt{{2\over N} } \int \left[\prod_{i=1}^6 dU_i\right] \; V(U_1,U_2;U_3,U_4;U_5,U_6)\cr
&\times\left[ \widetilde{A}_2(X_1,X_2;U_1,U_2)\widetilde{A}_2(U_3,U_4;U_5,U_6)\widetilde{A}_2(X_3,X_4;U_5,U_6)\right.\cr
&\qquad + \widetilde{A}_2(X_3,X_4;U_1,U_2)\widetilde{A}_2(U_3,U_4;U_5,U_6)\widetilde{A}_2(X_5,X_6;X_1,X_2)\cr
& \qquad +\left. \widetilde{A}_2(X_5,X_6;U_1,U_2)\widetilde{A}_2(U_3,U_4;U_5,U_6)\widetilde{A}_2(X_1,X_2;X_3,X_4)\right]\cr
&-2\sqrt{{2\over N} } \int \left[\prod_{i=1}^6 dU_i\right]  \; V(U_1,U_2;U_3,U_4;U_5,U_6)\cr
&\underbrace{\qquad \times \widetilde{A}_2(X_1,X_2;U_1,U_2)\widetilde{A}_2(X_3,X_4;U_3,U_4)\widetilde{A}_2(X_5,X_6;U_5,U_6)}_{\equiv \widetilde{A}_3}
\end{align}
One can see that the first three terms are disconnected propagators in the point of view of bi-local fluctuation. The last term is a connected propagators.
Setting
\begin{equation}
X_1=(a,t_1)\;,\;X_2=(a,t_2)\;,\;X_3=(b,t_3)\;,\;X_4=(b,t_4)\;,\;X_5=(c,t_5)\;,\;X_6=(c,t_6).
\end{equation}
the (connected) three-point function of diagonal fluctuation becomes
\begin{align}
\widetilde{A}^{abc}_3&(t_1,t_2;t_3,t_4;t_5,t_6)\equiv -2\sqrt{{2\over N} } \int \left[\prod_{i=1}^6 du_i\right]  \cr
&\qquad\qquad\times V^{abc}(u_1,\cdots,u_6) \widetilde{A}^{aa}_2(t_1,t_2;u_1,u_2)\widetilde{A}^{bb}_2(t_3,t_4;u_3,u_4)\widetilde{A}^{cc}_2(t_5,t_6;u_5,u_6)
\end{align}
As we did in the Section~\ref{sec:two-point function}, we will amputate the external legs.
\begin{equation}
\widetilde{A}^{abc}_3(t_1,t_2;t_3,t_4;t_5,t_6)\equiv \int \left[\prod_{i=1}^6 du_i \;\psi_0(t_i,u_i) \right]A^{abc}_3(u_1,u_2;u_3,u_4;u_5,u_6)
\end{equation}
It is also convenient to define amputated cubic vertex
\begin{align}
&\widetilde{V}^{abc}_3(t_1,t_2;t_3,t_4;t_5,t_6)\cr
&\equiv \int \left[\prod_{i=1}^6 du_i \;\psi_0(t_i,u_i) \right]V^{abc}_3(u_1,u_2;u_3,u_4;u_5,u_6)\cr
&=-\delta_{a,b}\delta_{a,c}{1 \over 8 }\left[\psi_0(t_6,t_1)\psi_0(t_2,t_3)\psi_0(t_4,t_5)+(\mbox{anti-sym})\right]\cr
&-\delta_{a,b}\delta_{a,c} 3J^2\int du_1du_2\; {1\over 4}\left[\psi_0(u_1,u_2)\psi_0(t_1,u_1)\psi_0(t_2,u_2)\psi_0(t_3,u_1)\psi_0(t_4,u_2)\psi_0(t_5,u_1)\psi_0(t_6,u_2)\right.\cr
&\qquad\qquad\qquad\qquad\qquad\qquad\left.+(\mbox{anti-sym})\right]
\end{align}
As we did for the two-point function to compare with \cite{Polchinski:2016xgd}, we picked up connected propagator $\Gamma^{ab}_2$ in the point of view of the fermion.
%
%
Using $\Gamma^{ab}_2$, the three-point function $A^{abc}_3$ can be also separated into connected and disconnected ones with respect to the fermion.
\begin{eqnarray}
&&A^{abc}_3(t_1,t_2;t_3,t_4;t_5,t_6)\cr
&\equiv& -\sqrt{{8\over N} } \int \left[\prod_{i=1}^6 du_i\right]  \; \widetilde{V}^{abc}(u_1,\cdots,u_6) A^{aa}_2(t_1,t_2;u_1,u_2)A^{bb}_2(t_3,t_4;u_3,u_4)A^{cc}_2(t_5,t_6;u_5,u_6)\cr
&=&\delta_{a,b}\delta_{a,c}(\mbox{disconnected diagrams})+\delta_{a,b}\delta_{a,c}\Gamma_3(t_1,t_2;t_3,t_4;t_5,t_6)
\end{eqnarray}
Noting that the two-point function of diagonal fluctuations is
\begin{equation}
\Gamma^{ab}_2(t_1,t_2;t_3,t_4)=\delta_{a,b}\Gamma_2(t_1,t_2;t_3,t_4),
\end{equation}
the connected 6-point function (in the point of view of the fermion) can be written as follows.
%
%
\begin{align}
&\sqrt{N\over 8}\Gamma_3(t_1,t_2;t_3,t_4;t_5,t_6)\cr
=&\int du_1du_2\; {1\over 2} \left[\Gamma_2(t_3,t_4;t_2,u_1)\psi_0(u_1,u_2) \Gamma_2(t_5,t_6;u_2,t_1) - (t_1 \; \longleftrightarrow\; t_2) \right]\cr
&\qquad+(\mbox{cyclic in }(t_1,t_2), (t_3,t_4)\mbox{ and } (t_5,t_6) )\cr
&+\int \left[\prod_{i=1}^6du_i\right]\; \Gamma_2(t_1,t_2;u_1,u_2)\psi_0(u_2,u_3) \Gamma_2(t_3,t_4;u_3,u_4) \psi_0(u_4,u_5)\Gamma_2(t_5,t_6;u_5,u_6)\psi_0(u_6,u_1)\cr
&-3J^2 {1\over 4}\left[\psi_0(t_1,t_2)\delta(t_3-t_1)\delta(t_5-t_1)\delta(t_4-t_2)\delta(t_6-t_2)+(\mbox{anti-sym})\right]\cr
&+3J^2 \int du_1 du_2\; {1\over 2} \left[\Gamma_2(t_1,t_2;u_1,u_2)\psi_0(u_1,t_3)\psi_0(u_2,t_4)\delta(t_3-t_5)\delta(t_4-t_6)\psi_0(t_3,t_4)-(t_3\; \longleftrightarrow\; t_4)\right]\cr
&\qquad+(\mbox{cyclic in }(t_1,t_2), (t_3,t_4)\mbox{ and } (t_5,t_6) )\cr
&-3J^2 \int du_3du_4du_5du_6\; \psi_0(t_1,t_2)\psi_0(t_1,u_3)\psi_0(t_2,u_4)\Gamma_2(t_3,t_4;u_3,u_4)\times\psi_0(t_1,u_5)\psi_0(t_2,u_6)\Gamma_2(t_5,t_6;u_5,u_6)\cr
&\qquad-(\mbox{cyclic in }(t_1,t_2), (t_3,t_4)\mbox{ and } (t_5,t_6) )\cr
&+3J^2\int \left[\prod_{i=1}^6 du_i\right]dv_1dv_2\; \psi_0(v_1,v_2) \Gamma_2(t_1,t_2;u_1,u_2)\psi_0(u_1,v_1)\psi_0(u_2,v_2)\cr
&\qquad\qquad\times\Gamma_2(t_3,t_4;u_3,u_4)\psi_0(u_3,v_1)\psi_0(u_4,v_2)\Gamma_2(t_5,t_6;u_5,u_6)\psi_0(u_5,v_1)\psi_0(u_6,v_2)
\end{align}
This is represented in the figure~\ref{fig:three-point function}.

\section{Evaluation of the Time Coordinate Action}
\label{app:evaluation of the time action}
In this Appendix, we consider the evaluation of the action for the collective time coordinate $f(t)$:
	\begin{align}
		S[f] \, = \, \frac{N}{2} \int dt_1 \, \Big[ \partial_1 \Psi_{0, f}(t_1, t_2)\Big]_{t_2 \to t_1} \, .
	\end{align}
Considering the generalized case of order $q$ fermionic coupling \cite{Kitaev:2015, Maldacena:2016hyu}
(which simply corresponds to changing the power for in the bi-local action of Eq.(\ref{eq:collective action})) the background solution is given by
	\begin{align}
		\Psi_{0, f}(t_1, t_2) \, = \, b \left( \frac{\sqrt{|f'(t_1)f'(t_2)|}}{|f(t_1) -f(t_2)|} \right)^{\frac{2}{q}} \, .
	\end{align}
where
	\begin{align}
		 b \, = \, - \, \left[ \frac{\tan\left( \frac{\pi}{q} \right)}{2\pi} \left( 1- \frac{2}{q} \right) \right]^{\frac{1}{q}} \, ,
	\end{align}
With $q=2+2\varepsilon$, we can think of an $\varepsilon$ expansion where the leading approximation corresponding to
the $q=2$ case, which is easily evaluated as follow:

One expands the denominator of the critical solution in the $t_2 \to t_1$ limit as
	\begin{align}
		\frac{1}{|f(t_1) -f(t_2)|}
		\, = \, \frac{1}{|f'(t_2)||t_1 -t_2|} \, - \, \frac{|f''(t_2)|}{2|f'(t_2)|^2} \, + \, \frac{|f''(t_2)|^2}{4|f'(t_2)|^3} \, |t_1-t_2| \, - \, \frac{|f'''(t_2)|}{6|f'(t_2)|^2} \, |t_1-t_2| \, + \, \cdots \, .
	\end{align}
We also need to expand the $\sqrt{|f'(t_1)|}$ in the numerator as
	\begin{align}
		\sqrt{|f'(t_1)|}
		\, = \, |f'(t_2)|^{\frac{1}{2}} \left[\, 1 \, + \, \frac{|f''(t_2)|}{2|f'(t_2)|} \, |t_1-t_2| \, - \, \frac{|f''(t_2)|^2}{8|f'(t_2)|^2}\, |t_1-t_2|^2 \, + \, \frac{|f'''(t_2)|}{4|f'(t_2)|}\, |t_1-t_2|^2 \, + \, \cdots \right] \, .
	\end{align}
Therefore, now we have
	\begin{align}
		\Psi_{0, f}(t_1, t_2) \, = \, - \, \frac{1}{\pi J}
		\left[ \frac{1}{|t_1-t_2|} \, + \, \frac{1}{12} \, \frac{|f'''(t_2)|}{|f'(t_2)|} \, |t_1-t_2| \, - \, \frac{1}{8} \, \frac{|f''(t_2)|^2}{|f'(t_2)|^2} \, |t_1-t_2| \, + \, \cdots \right] \, .
	\end{align}
In the following, we eliminate the first term, which gives a divergence.
Therefore, taking the derivative and the limit, the action is given by
	\begin{align}
		S[f] \, &= \, - \, \frac{N}{24\pi J} \int dt_1 \, \left[ \, \frac{f'''(t_1)}{f'(t_1)} \, - \, \frac{3}{2} \left( \frac{f''(t_1)}{f'(t_1)} \right)^2 \, \right] \nonumber\\
	\end{align}
The correcting terms in epsilon expansion are expected to always be of the Schwarzian derivative form, correcting only the overall factor in front of the Lagrangian.
These and higher order evaluations due to $J$-corrections are left for future publication.

\section{Identity}
\label{app:identity}
In this appendix, we will prove that
\begin{equation}
I(t_1,t_2)\equiv\int {dt_3dt_4\over |t_3-t_4|}\; \psi_0(t_1,t_3)\psi_0(t_2,t_4) u_{\nu w}(t_3,t_4)= {8\pi^{1\over 2} \over J\nu} \tan\left({\pi \nu \over 2}\right) u_{\nu w}(t_1,t_2).
\end{equation}
For $J\gg 1/|t|$, the background field reads
\begin{equation}
\psi_0(t_1,t_2)=\psi_0(t_1-t_2)=- {\sgn(t_1-t_2) \over (4\pi )^{1/4} \sqrt{J|t_1-t_2|}}  \, .
\end{equation}
Fourier transformation of the background field into momentum space 
\begin{equation}
\psi_0(w)=\int_{-\infty}^\infty dt \; \Psi_0(t) e^{iwt} 
={ \pi^{1/4} \; \sgn (w)\over  i \sqrt{J |w|} }
\end{equation}
Hence, the inverse\footnote{when the bi-local field is considered as a matrix in bi-local} of the background field can be written as
\begin{equation}
\psi_0^{-1}(t_1,t_2)={1\over 2\pi} \int dp \;{i\sqrt{J|p|}\sgn(p) \over \pi^{1/4} } e^{ip(t_1-t_2)}
\end{equation}
It is convenient to transform bi-local space $(t_1,t_2)$ to $(\tau,z)$ in \eqref{eq:transformation}:
\begin{align}
\tau_1={1\over 2}(t_1+t_2)\qquad,\qquad \tau_2={1\over 2}(t_3+t_4)\\
z_1={1\over 2}(t_1-t_2)\qquad,\qquad z_2={1\over 2}(t_3-t_4)
\end{align}
%
%
%
Then, one can show that 
\begin{equation}
I(t_1,t_2)
= - { 2  e^{-iw\tau_1} \over \pi^{1/2} J} \int {dpdz_2\over z_2}\; {\sgn(p)\sgn(w-p)\over \sqrt{|p| |w-p|} }e^{i(2p-w)(z_2-z_1)}   Z_\nu(|wz_2|)\label{eq:integration identity 1}
\end{equation}
Using the integral representations of Bessel function
\begin{align}
J_0(x)&= {2\over \pi} \int_1^\infty dt {\sin (xt) \over \sqrt{t^2-1}}\qquad (x>0)\\
Y_0(x)&= -{2\over \pi} \int_1^\infty dt {\cos( xt) \over \sqrt{t^2-1}}\\
J_0(x)&= {2\over\pi}\int_0^1 dt {\cos xt\over \sqrt{1-t^2} }\;,
\end{align}
the integral with respect to $p$ in \eqref{eq:integration identity 1} can be evaluated as follows.
\begin{equation}
\int dp\;  {\sgn(p)\sgn(w-p)\over \sqrt{|p||w-p|}} e^{i(2p-w)z}
=2\pi [J_0(wz)+Y_0(wz)]
\end{equation}
Then, we have
\begin{equation}
I(t_1,t_2)
=- {  4 \pi^{1/2}e^{-iw\tau_1} \sgn(w)\over J } \int  {du_2\over u_2} \left[ J_0(u_1-u_2))+Y_0(u_1-u_2)\right]Z_\nu(|u_2|)
\end{equation}
where $u_i\equiv wz_i$ ($i=1,2$). Using another integral representations of the Bessel functions
\begin{align}
J_0(x)=&{2\over \pi}\int_0^\infty dt\; \sin(x\cosh t) \\
Y_0(x)=&-{2\over \pi}\int_0^\infty dt\; \cos(x\cosh t) \; ,
\end{align}
one can show that for $\nu={3\over 2}+2n$ $(n=0,1,2,\cdots)$ and $\nu=ir$ $(r>0)$
\begin{equation}
 \int du_2  \;  [J_0(u_1-u_2)+Y_0(u_1-u_2)] {Z_{\nu_2}(|u_2|) \over u_2 } =-{2\sgn(u_1)\over \nu} \tan{\pi \nu \over 2}Z_\nu(|u_1|).
\end{equation}
Therefore, we have the identity 
\begin{equation}
I(t_1,t_2)= {  8 \pi^{1\over 2}\tan\left({\pi \nu \over 2}\right) \over J \nu }  e^{-iw\tau_1} \sgn(z_1) Z_{\nu}(|u_1|)=-{16\pi^{1\over 2} \over 3J}g(\nu) u_{\nu w}(\tau_1,z_1)
\end{equation}
where we used $g(\nu)$ defined in~\cite{Polchinski:2016xgd}
\begin{equation}
g(\nu)=-{3\over 2\nu }\tan\left({\pi \nu \over 2}\right)
\end{equation}
%
%


\end{document}